\documentclass[12pt,preprint]{aastex}

\usepackage{bm}
\shorttitle{GS and SOR methods for radiative transfer with PRD}
\shortauthors{Sampoorna \& Trujillo Bueno}
\begin{document}
\title{Gauss-Seidel and Successive Overrelaxation Methods for 
Radiative Transfer with Partial Frequency Redistribution}
\author{M. Sampoorna$^1$ and J. Trujillo Bueno$^{1,2,3}$} 
\affil{$^1$Instituto de Astrof\'isica de Canarias,
E-38205 La Laguna, Tenerife, Spain}
\affil{$^2$Departamento de Astrof\'isica, Facultad de F\'isica, Universidad de La Laguna, 
Tenerife, Spain}
\affil{$^3$Consejo Superior de Investigaciones Cient\'ificas, Spain}
\affil{{\bf{Accepted in February 2010 for publication in The Astrophysical Journal}}}

\begin{abstract}
The linearly-polarized solar limb spectrum that is produced by
scattering processes contains a wealth of information
on the physical conditions and magnetic fields of the solar outer atmosphere, 
but the modeling of many of its strongest spectral lines requires solving an 
involved non-LTE radiative transfer problem accounting for partial 
redistribution (PRD) effects. Fast radiative transfer methods for the numerical solution 
of PRD problems are also needed for a proper treatment of hydrogen lines when aiming at 
realistic time-dependent magnetohydrodynamic simulations of the solar chromosphere.
Here we show how the two-level atom PRD problem 
with and without polarization can be solved accurately and efficiently via the 
application of highly convergent iterative schemes based on the Gauss-Seidel 
(GS) and Successive Overrelaxation (SOR) radiative transfer methods that had 
been previously developed for the complete redistribution (CRD) case. Of 
particular interest is the Symmetric SOR method, which allows us to reach the 
fully converged solution with an order of magnitude of improvement in the 
total computational time with respect to the Jacobi-based local ALI 
(Accelerated Lambda Iteration) method.
\end{abstract}

\section{Introduction}

The phenomenon of scattering in a spectral line is a complicated physical process where 
partial correlations between the incoming and outgoing photons can occur 
\citep[e.g.,][]{mih78,can85,os94}. This happens when the shape of the incident spectrum that populates the upper level via radiative absorptions is not flat over the line. These partial redistribution (PRD) effects tend to be more important in strong 
lines, such as the solar Mg {\sc ii} and Ca {\sc ii} resonance lines and Lyman $\alpha$.
In particular, the wings of the intensity profiles of these lines are strongly affected by PRD effects, especially 
concerning observations close to the edge of the solar disk. 

Only a small number of solar spectral lines show conspicuous PRD signatures 
in their emergent {\em intensity} profiles. However, a substantially larger fraction show clear hints of PRD effects 
in the fractional linear polarization $Q/I$ profiles that result from scattering processes 
in quiet regions of the solar atmosphere (e.g., see the classification proposed by Belluzzi \& Landi Degl'Innocenti 2009 of the various $Q/I$ shapes found in the linearly-polarized solar limb spectrum observed by Stenflo \& Keller 1997 and by Gandorfer 2000, 2002, 2005). To achieve a rigorous modeling of the weak polarization signals that constitute this so-called {\em Second Solar Spectrum} is very important, mainly because it contains valuable information on the magnetism of the extended solar atmosphere \citep[e.g., the review by][]{jtb09}. To this end it is crucial to solve accurately and efficiently the non-LTE 
(Local Thermodynamic Equilibrium) radiative transfer problem of resonance line polarization taking into account PRD effects. The present paper represents a contribution towards this goal.

Fast iterative methods based on operator splitting were introduced 
to astrophysics by \citet{can73} for unpolarized radiative transfer with 
complete redistribution (CRD) in scattering. 
Extensions of this type of methods to PRD were done by 
\citet{varandcan76}, and later by \citet{sch83} and \citet{rh01}. These methods 
are widely known today as Accelerated Lambda Iteration (ALI) methods 
\citep[e.g., the review by][]{hubeny03}. An 
optimum choice for the approximate lambda operator is the diagonal of 
the true lambda operator, which was introduced in the seminal paper 
by \citet{olsetal86}. This Jacobi based method for solving the two-level 
atom problem with CRD was generalized by 
\citet[][hereafter PA95]{palandaue95} to unpolarized PRD radiative transfer. 

Superior radiative transfer methods based on Gauss-Seidel (GS) and 
successive overrelaxation (SOR) iteration were developed by 
\citet[][hereafter TF95]{jtbandfb95}. The convergence rate of these 
iterative schemes is equivalent to that corresponding to upper or lower triangular approximate lambda 
operators, but without the need of constructing and inverting such operators. 
Therefore, the computing time per iteration is similar to that of the 
Jacobi scheme or local ALI method, but the number of iterations needed to 
reach convergence is an order of magnitude smaller. In their paper, TF95 
suggested the strategy to generalize their GS-based methods to the multilevel atom case.
\citet*{fbpetal97} implemented such a MUltilevel GAuss Seidel  
method (MUGA; see also Fabiani Bendicho \& Trujillo Bueno 1999 and Asensio Ramos \& Trujillo Bueno 
2006 for its generalization to 3D and spherical geometries) and combined it with a non-linear 
multigrid iterative scheme to produce multilevel radiative transfer programs whose convergence rates are  
insensitive to the grid size. Here we present the generalization of the GS and SOR radiative transfer methods
of TF95 to the two-level atom PRD problem, with and without scattering 
polarization.

An alternative iterative scheme for solving radiative transfer problems is the 
preconditioned bi-conjugate gradient method \citep{cas04}, which has 
been recently applied to plane-parallel \citep{pfanda09,knnetal09} and 
spherical geometries \citep{lsaetal09}. 
Its rate of convergence is similar to that of the optimal Symmetric 
SOR method of TF95 \citep[see][]{cas04}, but an efficient generalization of the 
preconditioned bi-conjugate gradient method to the multilevel atom case 
is presently an unsolved problem.

A suitable generalization of the local ALI method to the Zeeman line transfer
problem was done by \citet{jtbandeld96}. Extensions of 
the Jacobi, GS, and SOR schemes to scattering polarization 
were carried out by \citet[][CRD Jacobi]{fauetal97}, 
\citet[][PRD Jacobi]{fpmf97}, and \citet[][CRD Jacobi, GS and SOR]{jtbrms99}. 
All these iterative schemes for non-LTE 
radiative transfer were generalized to solve CRD multi-level scattering 
polarization problems in the presence of a magnetic field, including the 
possibility of atomic polarization in all levels (Trujillo Bueno 1999; Manso 
Sainz \& Trujillo Bueno 2003, see also Trujillo Bueno 2003). 
%\citep{jtb99,rmsjtb03}.
However only the Jacobi iterative scheme has been applied to solve the two-level 
atom polarized PRD radiative transfer problem in the presence of an external magnetic 
field (Nagendra et al. 1999; Fluri et al. 2003; Sampoorna et al. 2008, see also 
the reviews by Nagendra 2003; Nagendra \& Sampoorna 2009). 
%\citep{knnetal99,fluetal03,sametal08}. 
Here we extend the GS and SOR iterative methods to solve 
polarized PRD problems in the absence or in the presence of magnetic fields 
which do not break the axial symmetry of the problem (e.g., the micro-turbulent 
field case).

The accuracy of any iterative method for a given 
depth grid resolution is determined by the truncation error or the true 
error \citep*[see][]{aueetal94}. So far the study of the true error is 
limited to only CRD problems (e.g., Auer et al. 1994; TF95; 
Faurobert-Scholl et al. 1997; Trujillo Bueno \& Manso Sainz 1999; 
Chevallier et al. 2003). Hence, in this paper we discuss certain aspects of the 
true error for PRD problems. 

For our study we consider all the three angle-averaged (AA)
redistribution functions of \citet{hum62}, namely $R_{\rm I,II,III, AA}$, 
and the linear combination of $R_{\rm II,AA}$ and $R_{\rm III, AA}$. 
We recall that physically (1) $R_{\rm I,AA}$ represents the case 
of infinitely sharp lower and upper levels (or pure Doppler redistribution 
in the laboratory frame); 
(2) $R_{\rm II,AA}$ represents the case of infinitely sharp lower level 
and radiatively broadened upper level (coherent scattering in the atomic 
frame); (3) $R_{\rm III,AA}$ represents 
the case of infinitely sharp lower level and radiatively as well as 
collisionally broadened upper level (CRD in the atomic frame). 

The logical structure of this paper is the following\,: 
\S\S~2, 3, and 4 are devoted to unpolarized PRD radiative transfer. 
We first recall the basic equations, the Jacobi scheme used by 
PA95 for $R_{\rm II,AA}$ redistribution, and then present 
briefly the extension of this scheme to $R_{\rm I,III,AA}$ redistributions. 
Next, we present the generalization of the GS and SOR schemes
of TF95 to PRD. A detailed study of the true error for all the 
three iterative schemes with AA redistribution functions 
is conducted in \S~4. \S\S~5, 6, and 7 are devoted to polarized 
PRD radiative transfer. In \S~5 we present the basic 
equations of polarized radiative transfer. Our generalization of the 
Jacobi, GS, and SOR schemes of \S~3 to scattering polarization is presented in 
\S~6. A detailed study of the true error for polarized PRD case is given in 
\S~7. Concluding remarks are given in \S~8. 

\section{Unpolarized PRD radiative transfer}
We consider the case of scattering on a two-level atom with a 
background continuum. The scattering mechanism is described by the 
AA redistribution functions of \citet{hum62}. Furthermore, we approximate the 
stellar atmosphere by a one-dimensional plane parallel slab of total optical 
thickness $T$. Under these assumptions, the scalar radiative transfer equation 
is given by 
\begin{equation}
{{\rm d}  \over {\rm d} \tau} I_{x\mu}(\tau)=I_{x\mu}(\tau) - S_x(\tau), 
\label{srte}
\end{equation}
where $I_{x\mu}(\tau)$ is the specific intensity, $x$ is the 
frequency from line center in units of the Doppler width, 
$\mu= \cos \theta$, with $\theta$ being the angle between the 
ray and the atmospheric normal, and the optical depth $\tau$ is defined 
by ${\rm d}\tau= -(\chi_l\phi_x+\chi_c){\rm d}z/\mu$, with $\phi_x$ the 
normalized Voigt profile function, $\chi_c$ and $\chi_l$ the 
continuum and line opacities, 
and $z$ the distance along the normal to the atmosphere. Hereafter, we omit 
the $\tau$ dependence of the intensity and source function 
for notational simplicity. The monochromatic source function is given by 
\begin{equation}
S_x = {\phi_x S_{lx} + r B\over \phi_x+r} ,
\label{source_total}
\end{equation}
where $r=\chi_c/\chi_l$, and $B$ is the Planck function at the line 
frequency. The line source 
function is given by
\begin{equation}
S_{lx} = (1-\epsilon) \bar J_x + \epsilon B,
\label{linesource}
\end{equation}
where $\epsilon$ is the collisional destruction probability. The PRD scattering 
integral or mean PRD intensity is given by
\begin{equation}
\bar J_x = \int g^k_{xx^\prime} J_{x^\prime} {\rm d}x^\prime,
\label{scat_int}
\end{equation}
with the mean intensity
\begin{equation}
J_{x} = {1 \over 2} \int I_{x\mu} {\rm d}\mu\,.
\label{mean_int}
\end{equation}
In Equation~(\ref{scat_int}), $g^k_{xx^\prime}=R_{k,{\rm AA}}(x,x^\prime)/\phi_x$, 
where $R_{k, {\rm AA}}$ with $k={\rm I,\,II,\ and\ III}$ are the 
AA redistribution functions of \citet{hum62}. Their functional 
as well as the graphical form can be found in \citet{hum62}, \citet{mih78} 
and \citet{hei81}. 

\section{Iterative methods for unpolarized PRD radiative transfer}
We write the formal solution of the radiative 
transfer equation~(\ref{srte}) as
\begin{equation}
I_{x\mu} = \Lambda_{x\mu}[S_x]+T_{x\mu}\ ,
\label{formal_solution}
\end{equation}
where $T_{x\mu}$ gives the transmitted specific intensity due to the 
incident radiation at the boundary and $\Lambda_{x\mu}$ is a $N
\times N$ operator whose elements depend on the optical distances 
between the grid points, with $N$ being the number of spatial grid points. 

A suitable formal solution method for the numerical solution of 
Equation~(\ref{srte}) is the short-characteristics method 
\citep{kunandaue88,aueandpal94,aueetal94}. This method is based on the 
assumption that the variation of the source function with the optical depth 
along the ray under consideration is a parabola between three consecutive grid
points. Thus, if $M$ represents an upwind point, $O$ represents the point of 
interest and $P$ the downwind point, then 
the intensity at point $O$ is given by 
\begin{eqnarray}
I_{x\mu,O} = I_{x\mu,M} {\rm e}^{-\Delta \tau_{M}} 
+ \Psi_{x,M}(\mu) S_{x,M} \nonumber \\
+ \Psi_{x,O}(\mu) S_{x,O} + \Psi_{x,P}(\mu) S_{x,P},
\label{sc_i}
\end{eqnarray}
where $\Delta \tau_{M}$ is the optical distance on segment $MO$, 
$\Psi_{x,M,O,P}$ are functions of the optical distances between $O$ and  
$M$ and between $O$ and $P$, and 
$S_{x,M,O,P}$ are source function values at the points $M$, $O$, and $P$, 
respectively. However, at the boundaries a linear interpolation for the source 
function along the points $M$ and $O$ is used for the rays going out of such 
boundaries. 

Following \citet{jtb03} we write the mean PRD intensity at the spatial grid 
point `$i$' as
%\begin{eqnarray}
%J_{x,i}&=&\Lambda_{x,i1}S^{a}_{x,1} +\cdots +\Lambda_{x,ii-1}S^{a}_{x,i-1} \nonumber\\
%&&+\Lambda_{x,ii}S^{b}_{x,i} +\Lambda_{x,ii+1}S^{c}_{x,i+1} + 
%\cdots \nonumber \\
%&&+\Lambda_{x,iN}S^{c}_{x,N} + T_{x,i}\ ,
%\label{mean_int_iter}
%\end{eqnarray}
\begin{eqnarray}
\bar J_{x,i}&=&\int g^k_{xx^\prime} \bigl[\Lambda_{x^\prime,i1}S^{a}_{x^\prime,1} +
\cdots +\Lambda_{x^\prime,ii-1}S^{a}_{x^\prime,i-1} \nonumber\\ &&+
\Lambda_{x^\prime,ii}S^{b}_{x^\prime,i} +\Lambda_{x^\prime,ii+1}
S^{c}_{x^\prime,i+1} +
\cdots \nonumber \\ &&+\Lambda_{x^\prime,iN}S^{c}_{x^\prime,N}\bigr]{\rm d}x^\prime 
+ {\bar T}_{x,i}\,, 
\label{scat_int_iter}
\end{eqnarray}
where $\bar T_{x,i}$ is given by Equation~(\ref{scat_int}) but with 
$J_{x^\prime}$ replaced by $T_{x^\prime}$ (which is given by Equation~(\ref{mean_int}) 
with $I_{x^\prime\mu}$ replaced by $T_{x^\prime\mu}$). For each frequency $x$, 
$\Lambda_{x,ii^\prime}$ is obtained 
by integrating $\Lambda_{x\mu}$ over all the directions $\mu$ 
of the incoming and outgoing radiation beams of the angular quadrature 
chosen for the numerical integration. Furthermore, $a,b$ and $c$ are 
simply symbols that will be useful to indicate weather we choose the `old' 
or the `new' values of the source function. 
In the following subsections we, 
first, briefly recall the Jacobi iterative method, and then present the GS 
and SOR iterative schemes. 

\subsection{Jacobi iterative scheme}
Let us recall first the Jacobi iterative scheme presented in PA95. 
This scheme is obtained by choosing in Equation~(\ref{scat_int_iter}) 
${a}={c}={\rm old}$, but ${b}={\rm new}$, which gives\,:
\begin{equation}
\bar J_{x,i}
%&=& \bar J^{\rm old}_{x,i} + \int g^k_{xx^\prime} 
%\Lambda_{x^\prime,ii}[S^{\rm new}_{x^\prime,i}-S^{\rm old}_{x^\prime,i}]
%{\rm d}x^\prime, \nonumber \\
= \bar J^{\rm old}_{x,i}  + \int g^k_{xx^\prime}
\Lambda_{x^\prime,ii}\,p_{x^\prime}\,\delta S_{lx^\prime,i}\,{\rm d}x^\prime\,,
\label{scat_int_jacobi}
\end{equation}
where we  have used Equation~(\ref{source_total}). In the above equation 
$p_x = \phi_x/(\phi_x+r)$ and 
$\delta S_{lx,i}=S^{\rm new}_{lx,i} - S^{\rm old}_{lx,i}$. 
Using Equation~(\ref{scat_int_jacobi}) in Equation~(\ref{linesource}), we 
obtain the following expressions for the line source function corrections\,:
\begin{eqnarray}
&&\delta S_{lx,i} - (1-\epsilon)\int g^k_{xx^\prime}\,
\Lambda_{x^\prime,ii}\,p_{x^\prime}\delta S_{lx^\prime,i}\, {\rm d}x^\prime
\nonumber \\
&&= (1-\epsilon)\bar J^{\rm old}_{x,i} + \epsilon B - S^{\rm old}_{lx,i}\ .
\label{line_source_correction_jacobi}
\end{eqnarray}
Following PA95 we define a frequency dependent residual as
\begin{equation}
r_{x,i} = (1-\epsilon)\bar J^{\rm old}_{x,i} + \epsilon B - S^{\rm old}_{lx,i}\ .
\label{residual}
\end{equation}

Thus at each depth point we have to solve $N_x$ linear equations, 
where $N_x$ is the number of frequency points. The simplest (but numerically 
expensive) way to find the solution 
of the system of linear equations~(\ref{line_source_correction_jacobi}) 
is by matrix inversion as follows\,:
\begin{equation}
\delta {\bm S} = {\bf A}^{-1}{\bm r},
\label{fbf}
\end{equation}
where at each depth point $i$, ${\bm r}$ is a vector of length $N_x$, and 
the matrix ${\bf A}$ is of 
dimension $N_x \times N_x$, and its elements are given by
\begin{equation}
A_{mn}=\delta_{mn}-(1-\epsilon)\tilde g^k_{mn}\,\Lambda_{n,ii} \,p_n\, ; 
\ \ m,n=1,\cdots,N_x\,.
\label{fbf_mat}
\end{equation}
Here $\delta_{mn}$ is the Kronecker's symbol, $\tilde g^k_{mn}$ are 
the redistribution weights, and the indices $m$ and $n$ 
refer respectively to the discretized values of $x$ and $x^\prime$. Note 
that for isothermal slabs the matrix ${\bf A}$ can be computed only once and 
can be inverted and stored. This method was referred to as the 
Frequency-by-Frequency (FBF) method by PA95, which was 
developed by the authors for type II redistribution. It is easy 
to note that the same method can be easily applied to the type I and type 
III redistributions and a linear combination of type II and type III 
redistributions without any further effort. 

The above FBF method involves the inversion of a matrix, which can be huge for 
realistic problems. For this reason, PA95 proposed a faster but equally 
robust method for the case of a type II redistribution function. 
In the following sub-section we briefly discuss this more efficient method, 
which is presented in more detail in PA95. 

\subsubsection{CRD-CS or Core-Wing method for type II redistribution}
It is well known that $g^{\rm II}_{xx^\prime}$ behaves like 
CRD in the line core and like coherent scattering 
in the wings \citep[see][]{mih78}. Taking advantage of this 
fact, to reduce the computational cost involved in the 
calculation of $\delta S_{lx,i}$, one can introduce a 
core-wing approximation to the true redistribution function 
$g^{\rm II}_{xx^\prime}$, namely 
\begin{eqnarray}
\label{crdcs}
g^{\rm II}_{xx^\prime} \approx 
\cases{\phi_{x^\prime}, &  for\ $x,x^\prime \le x_c$, \cr
\delta(x-x^\prime), & for\  $x^\prime > x_c$.}
\end{eqnarray}
Here $x_c$ is called the separation frequency that distinguishes 
between the line core and 
the wing. PA95 showed that $x_c=3.5$ Doppler widths is a physically 
reasonable choice and, hence, we adopt the same in this paper.  

Substituting Equation~(\ref{crdcs}) in 
Equation~(\ref{line_source_correction_jacobi}), 
the equation for $\delta S_{lx,i}$ takes the simpler form 
\begin{equation}
\delta S_{lx,i} = {r_{x,i} + (1-\alpha_x)\Delta T_i \over 1-\alpha_x
(1-\epsilon)p_x\Lambda_{x,ii}},
\label{line_source_correction_wing}
\end{equation}
where $\alpha_x$ are the splitting coefficients that allow for a smooth 
transition between the core and the wing. In the core $\alpha_x=0$ and in 
the wing $\alpha_x=g^{\rm II}_{xx}$. The frequency independent core 
integral $\Delta T_i$ is given by 
\begin{equation}
 \Delta T_i = (1-\epsilon)\int_{core} \phi_{x^\prime}p_{x^\prime}
\Lambda_{x^\prime,ii}\,\delta S_{lx^\prime,i}\, {\rm d}x^\prime.
\label{core_crd}
\end{equation}
To evaluate $\Delta T_i$, we consider only the core frequencies 
(i.e., $\alpha_x=0$) in Equation~(\ref{line_source_correction_wing}) 
and then apply the operator 
$(1-\epsilon)\int_{core} \phi_{x^\prime}p_{x^\prime}
\Lambda_{x^\prime,ii}[\,] {\rm d}x^\prime$. The resulting scalar 
equation for $ \Delta T_i$ can be easily solved to obtain 
$ \Delta T_i$ (see PA95 for more details). 
From Equation~(\ref{line_source_correction_wing}) we see that the wing 
frequencies drop out for $x \le x_c$, and in the wings 
the term multiplying $(1-\alpha_x)$ appears only as a frequency 
independent quantity. Thus, it is possible to find all the 
$\delta S_{lx,i}$ values from a simple scalar 
equation, thereby completely avoiding the solution 
of a system of equations irrespective of the number of frequency grid points. 

\subsubsection{Extending the CRD-CS method to type I and III redistribution}
The extension of the CRD-CS method to type III redistribution has been given 
in \citet{fluetal03}. To this end, the type III redistribution function is 
approximated by assuming CRD in the core and by setting it to zero in the 
wings. This is justified because the type III redistribution function does not 
show coherent peaks in the wings \citep[see][]{mih78}. 
However, we find that for the pure type III redistribution case we have 
to approximate $R_{\rm III,AA}$ by CRD throughout the frequency bandwidth 
to compute $\delta S_{lx,i}$. Setting it to zero in the 
wings leads to convergence problems. In the case of a linear combination of 
$R_{\rm II,AA}$ and $R_{\rm III,AA}$ (see \S~4.4, Equation~(\ref{r23_mix})), 
we find that as long as elastic collisions are small we can approximate 
$R_{\rm II,AA}$ by CRD-CS and $R_{\rm III,AA}$ by CRD in the line core and 
set it to zero in the wings. However, when elastic collisions are large we 
can approximate $R_{\rm II,AA}$ by CRD-CS, but $R_{\rm III,AA}$ should be 
approximated by CRD throughout the line profile, otherwise we have 
convergence problems. 

For soling the type I redistribution problem we approximate $R_{\rm I,AA}$ by 
CRD in the core and set it to zero in the wings for the computation of 
the $\delta S_{lx,i}$ corrections. 
This is justified as $R_{\rm I,AA}$ is a pure Doppler 
redistribution function and does not show coherent peaks in the wings 
\citep[see][]{mih78}. 

\subsection{Gauss-Seidel and SOR iterative schemes}
The radiative transfer methods based on GS and SOR iterations were developed 
by TF95 for the CRD case. In this section we extend such methods to solve 
unpolarized PRD problems. 

The GS iterative scheme is obtained by choosing ${c}={\rm old}$ and 
${a}={b}={\rm new}$ in Equation~(\ref{scat_int_iter}). This gives 
\begin{equation}
\bar J_{x,i} 
= \bar J^{\rm old+new}_{x,i} + \int g^k_{xx^\prime}
\Lambda_{x^\prime,ii}\,\delta S_{x^\prime,i}\,{\rm d}x^\prime,
\label{scat_int_gs}
\end{equation}
where $\delta S_{x,i}=p_x\,\delta S_{lx,i}$ and 
$\bar J^{\rm old+new}_{x,i} $ is the mean PRD intensity calculated 
using the `new' values of the source function at grid points $1,2,\cdots,i-1$ 
and the `old' values at points $i,i+1,\cdots,N$. The line 
source function corrections are now given by
\begin{eqnarray}
&&\delta S_{lx,i} - (1-\epsilon)\int g^k_{xx^\prime}
\Lambda_{x^\prime,ii}\,p_{x^\prime}\delta S_{lx^\prime,i}\, {\rm d}x^\prime
\nonumber \\ &&= (1-\epsilon)\bar J^{\rm old+new}_{x,i} + \epsilon B - S^{\rm old}_{lx,i}\ .
\label{line_source_correction_gs}
\end{eqnarray}
To compute $\delta S_{lx,i}$ we can apply either the FBF or the 
CRD-CS method discussed in \S~3.1. 
The SOR iterative scheme is 
obtained by doing the corrections as follows\,:
\begin{equation}
\delta S^{\rm SOR}_{lx,i}=\omega\delta S^{\rm GS}_{lx,i}\ ,
\label{line_source_correction_sor}
\end{equation}
where $\omega$ is a parameter with an optimum value between 1 and 2 which 
can be found using the method discussed in \S~2.4 of TF95. 
The Optimum value of $\omega$ is the one that 
leads to the highest rate of convergence. We find that the SOR method cannot 
be combined with the CRD-CS method for type II redistribution, while 
it works fine with the FBF method. The reason could be the way the wings 
are handled in the CRD-CS method. 

\subsubsection{The standard GS and SOR techniques}
It is worth noting that the GS iterative scheme is twice faster compared to 
the Jacobi scheme. A factor two of additional improvement can be achieved 
by implementing the symmetric GS scheme \citep[see TF95;][]{jtb03}. 
To explain the symmetric 
GS scheme, we first recall the GS scheme briefly, which is explained in 
greater detail in TF95. 

Following TF95, we consider two distinct parts\,: 
a incoming and outgoing. \\

\noindent
{\bf 1. Incoming part ($\mu<0$)\,:} One starts from the upper boundary 
($i=1$ with $i$ being the depth index), and 
determines the intensity of the incoming rays ($\mu < 0$) 
at all depths using the short-characteristics formal 
solver. Thus, at the end of the incoming section, one has calculated 
the incoming contribution to the mean PRD intensity, $\bar J_{x,i}(\mu < 0)$, 
at all the depth points ($i=1,\cdots N$). \\

\noindent
{\bf 2. Outgoing part ($\mu>0$)\,:} One now starts from the lower boundary  
($i=N$). Given that at this point the intensity is known, one can easily 
compute the total mean PRD intensity $\bar J_{x,N}$. We then use it to 
calculate $\delta S_{lx,N}$ and thereby the new source 
function $S^{\rm new}_{x,N}$ at the lower boundary. Now for the 
next depth point $i=N-1$, to calculate $I_{N-1}$ using Equation~(\ref{sc_i}) GS 
uses $S^{\rm new}_{x,N}$, $S^{\rm old}_{x,N-1}$ and $S^{\rm old}_{x,N-2}$. 
Then we compute the outgoing contribution to $\bar J_{x,N-1}(\mu > 0)$. 
However, note that the incoming contribution to $\bar J_{x,N-1}(\mu <0)$ was 
calculated with the old values of the source function, namely 
$S^{\rm old}_{x,N}$, $S^{\rm old}_{x,N-1}$ and $S^{\rm old}_{x,N-2}$. 
Therefore, to calculate the actual $\bar J^{\rm old+new}_{x,N-1}$ we 
add the following correction\,:
\begin{equation}
\Delta\bar J^{\rm in}_{x,N-1} = \int g^k_{xx^\prime} 
\Delta J^{\rm in}_{x^\prime,N-1} {\rm d}x^\prime,
\label{gs_correction_scat_int}
\end{equation}
where 
\begin{equation}
 \Delta J^{\rm in}_{x,N-1}=
{1 \over 2} \delta S_{x,N}\int_{-1}^{0} \Psi_{x,N}(\mu<0) {\rm d}\mu\,.
\label{gs_correction_mean_int}
\end{equation}
Here ``in'' stands for the incoming pass (see below). 
Once the actual $\bar J^{\rm old+new}_{x,N-1}$ 
is found one computes $\delta S_{lx,N-1}$ and 
the new source function $S^{\rm new}_{x,N-1}$. Because $S^{\rm new}_{x,N-1}$ 
is available now, before going to the next depth point the following 
correction should be added to the intensity $I_{x\mu>0}(N-1)$\,:
\begin{equation}
\Delta I^{\rm in}_{x\mu>0}(N-1) = \Psi_{x,N-1}(\mu>0)\delta S_{x,N-1}.
\label{gs_correction_int}
\end{equation}
The above procedure is then repeated for subsequent depth points. 
Clearly, unlike the Jacobi iterative scheme, the GS iterative method requires 
specific ordering of loops in the formal solver. The outermost loop is over 
the directions (first the incoming and then the outgoing rays). The next loop 
is over the spatial points, followed by the loop over different $|\mu|$ 
points. The innermost loop is over the frequencies. 

\subsubsection{The Symmetric GS and SOR techniques}
The symmetric GS iterative scheme discussed by \citet{jtb03} is obtained by 
introducing one more outer loop, which first does the GS iteration 
starting with the incoming ray (which we call 
incoming pass), and then the GS iteration starting with the outgoing 
ray (which we call outgoing pass). In the incoming pass all the GS steps 
are exactly the same as those described above. In the case of the outgoing 
pass, we again consider two parts to describe the symmetric GS case, namely 
the outgoing and the incoming 
parts. Let us clarify these parts\,: \\

\noindent
{\bf 3. Outgoing part of the outgoing pass ($\mu>0$)\,:} We start from the 
lower boundary ($i=N$) and compute the outgoing contribution to the 
$\bar J_{x,i}(\mu > 0)$ quantity at all depth points using the source 
function computed newly from the incoming pass. \\

\noindent
{\bf 4. Incoming part of the outgoing pass ($\mu<0$)\,:} We now start from 
the upper boundary. At the upper boundary ($i=1$), we can compute the new 
source function $S^{\rm new}_{x,1}$, as the intensity is known. For the next 
depth point $i=2$, we compute the intensity using $S^{\rm new}_{x,1}$, 
$S^{\rm old}_{x,2}$, and $S^{\rm old}_{x,3}$. Note that the so-called 
$S^{\rm old}_{x,i}$ for $i \ge 2$ are the new source functions obtained 
from the incoming pass (see above). Thus, we can now compute the incoming 
contribution to the $\bar J_{x,2}(\mu < 0)$. However, the outgoing contribution 
to $\bar J_{x,2}(\mu > 0)$ was calculated with the old values of the source 
function, namely, $S^{\rm old}_{x,1}$, $S^{\rm old}_{x,2}$, and 
$S^{\rm old}_{x,3}$. Therefore to calculate the actual 
$\bar J^{\rm old+new}_{x,2}$, one has to add the following correction
\begin{equation}
\Delta\bar J^{\rm out}_{x,2} = \int g^k_{xx^\prime} 
\Delta J^{\rm out}_{x^\prime,2}\, {\rm d}x^\prime,
\label{gssym_correction_scat_int}
\end{equation}
where
\begin{equation}
 \Delta J^{\rm out}_{x,2}=
{1 \over 2} \delta S_{x,1}\int_{0}^{+1} \Psi_{x,1}(\mu>0) {\rm d}\mu\,.
\label{gssym_correction_mean_int}
\end{equation}
In the above equations ``out'' denote the outgoing pass. 
Once the actual $\bar J^{\rm old+new}_{x,2}$ is found we can now compute the 
new source function $S^{\rm new}_{x,2}$. Since $S^{\rm new}_{x,2}$ is 
available, before going to the next depth point the following correction 
should be added to the intensity $I_{x\mu<0}(2)$\,:
\begin{equation}
\Delta I^{\rm out}_{x\mu<0}(2) = \Psi_{x,2}(\mu<0)\delta S_{x,2}.
\label{gssym_correction_int}
\end{equation}
The above procedure is then repeated for subsequent depth points.

This scheme together with the incoming pass (1 and 2) and outgoing pass 
(3 and 4) is nothing but the symmetric GS iterative scheme (hereafter SYM-GS). 
Thus, each call to the formal solver produces as an output two truly GS 
iterations. Clearly, the incoming pass has a convergence rate equivalent to 
that of a lower triangular approximate operator method and the outgoing 
pass has a convergence rate equivalent to that of an upper triangular 
approximate operator method (see TF95). This symmetric GS scheme 
can be extended to SOR also, which is then called Symmetric SOR 
\citep{jtb03}. The advantage of SSOR is that it is less 
sensitive to the choice of the optimum value of $\omega$ as compared to SOR 
\citep[see Fig.~1 of][]{jtb03}. Furthermore, unlike SOR, the SSOR method 
can be combined 
with standard acceleration techniques like Ng \citep[see][]{aue87,aue91} 
or orthomin's acceleration \citep{vin76,kleetal89,aue91}. 

\section{The true error of the numerical solutions}
Following \citet{aueetal94} we define three quantities that 
characterize any iterative scheme, namely (1) the maximum relative change 
$R_c$, (2) the maximum relative convergence error $C_e$, and (3) the maximum 
relative true error $T_e$. For a given level of grid resolution $g$ 
at the $n$th iterative stage these three quantities are defined as follows\,:
\begin{equation}
R_c(n,g)= \max_{\tau, x} \left[{|S_{lx}(n,g) - S_{lx}(n-1,g)|\over 
S_{lx}(n,g)}\right],
\label{mrc}
\end{equation}
\begin{equation}
C_e(n,g)= \max_{\tau, x} \left[{|S_{lx}(n,g) - S_{lx}(\infty,g)|
\over S_{lx}(\infty,g)}\right],
\label{mrce}
\end{equation}
\begin{equation}
T_e(n,g)= \max_{\tau, x} \left[{|S_{lx}(n,g) - S_{lx}(\infty,\infty)|
\over S_{lx}(\infty,\infty)}\right].
\label{te}
\end{equation}
In the above equations $(n=\infty,g)$ indicates that one is dealing with the 
fully converged solution on a grid resolution level $g$, while 
$(n=\infty,g=\infty)$ indicates 
the true solution on a grid of infinite resolution. $T_e(\infty,g)$ is nothing but 
the truncation error corresponding to a grid of finite resolution level $g$, 
and thus it determines the accuracy of the converged solution in that grid. 

In this paper we find the fully converged solution on a given grid resolution 
level $g$, by iterating until $R_c < 10^{-10}$. Beyond this value $R_c$ does 
not decrease any further, but simply fluctuates around it. 
The true solution required to calculate the true error is found by using a 
grid which is twice finer compared to the grid on which we seek the true 
error. 

Following TF95, in this paper we use the true error to 
determine the convergence properties of the iterative schemes. Here 
we show that the true error not only depends on the resolution of the spatial
grid and the accuracy of the formal solver, but also on the choice of the 
redistribution function. In the following subsections we discuss the true 
error separately for the $R_{\rm I,II,III,AA}$ functions and for cases with a 
linear combination of $R_{\rm II,AA}$ and $R_{\rm III,AA}$. 

\subsection{Pure Doppler redistribution - type I redistribution}
We recall that physically this case represents an atom with 
two sharp upper and lower levels. Thus, the line is infinitely sharp in the 
rest frame of the atom. In the laboratory frame it is broadened by the 
Doppler effect. This idealized case can hardly be applied to interpret any 
spectral lines, nevertheless it is an interesting academic case to study as 
it allows to examine the effects of pure Doppler redistribution by a 
Maxwellian velocity distribution. 

\begin{figure*}
%\centering  
%\includegraphics[height=5cm,width=16cm]{r1_tecemrc.eps}
\plotone{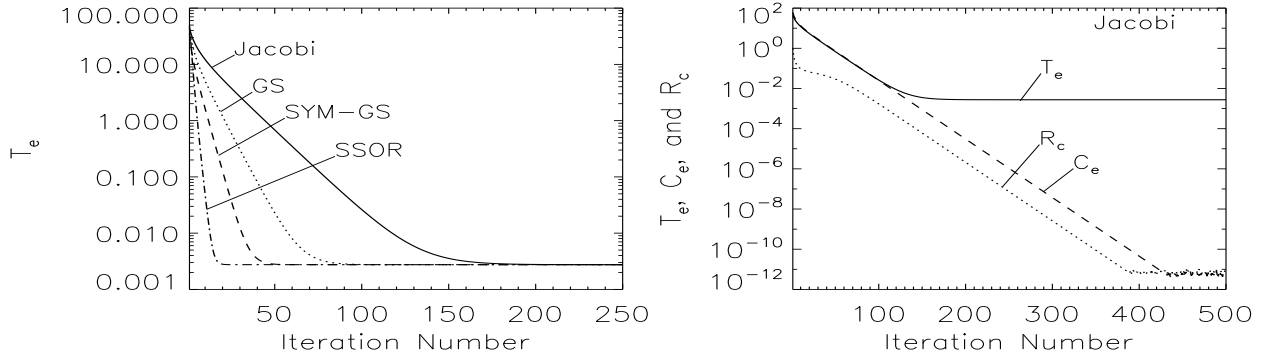}
\caption{
Pure Doppler (type I) redistribution.
Convergence properties of the various iterative schemes in a 
semi-infinite isothermal  
model atmosphere with $\epsilon=10^{-4}$ and a spatial grid 
with 9 points per decade ($\Delta Z=0.25$). Left panel\,: solid line (Jacobi), 
dotted line (GS), dashed line (SYM-GS), dot-dashed line (SSOR, 
$\omega_{\rm opt}=1.6$). Right panel\,: solid line ($T_e$), 
dotted line ($R_c$), and dashed line ($C_e$), are computed 
using the Jacobi iterative scheme. 
}
\label{r1_tercmrc}
\end{figure*}

Figure~\ref{r1_tercmrc} shows the convergence properties of different 
iterative schemes discussed in the previous section, applied here 
to type I redistribution. For all computations presented in this 
paper we consider the case of a semi-infinite atmosphere with 
the lower boundary condition $I_{x\mu}(\tau=T)=B$, and upper 
boundary condition $I_{x\mu}(\tau=0)=0$. 
The depth grid is constructed 
using the relation $\tau=\exp(-Z)$, where $Z=z/H$ (with $H$ the scale 
height), and $Z$ the height in units of $H$. We choose a uniform 
spacing of $\Delta Z$. 
For all the figures presented in this paper we have chosen 
$\Delta Z=0.25$ (which corresponds to 9 points per decade). 
A Gaussian quadrature with 3 inclinations $[0<\mu<1]$, 
and an equally spaced frequency grid with 41 points and a spacing 
of $0.25$ Doppler widths are used. Note that a frequency bandwidth 
of $0\le x\le 10$ is more than sufficient, as we are considering 
a pure Doppler redistribution (with zero damping). 
The collisional destruction probability $\epsilon=10^{-4}$. The Plank 
function $B$ is set to unity. Unless stated otherwise we set the continuum 
parameter $r$ to zero. From the left panel of Fig.~\ref{r1_tercmrc} we see that 
the convergence behavior of the different iterative schemes are exactly the 
same as that discussed in TF95. We note that for type I redistribution we 
obtain a true error of $2.7\times 10^{-3}$, while for the corresponding 
coherent scattering and CRD (with damping parameter $a=0$) cases we get a true 
error of $3.5\times 10^{-3}$, and $4.3\times 10^{-3}$, respectively. 

\subsection{Doppler, Natural and Collisional Broadening - type III redistribution}
Physically this case represents a resonance line with its upper level both 
radiatively and collisionally broadened. Collisions are 
so frequent that there is CRD in the rest frame of the atom. 

\begin{figure*}
%\centering  
%\includegraphics[height=5cm,width=16cm]{r3_tecemrc.eps}
\plotone{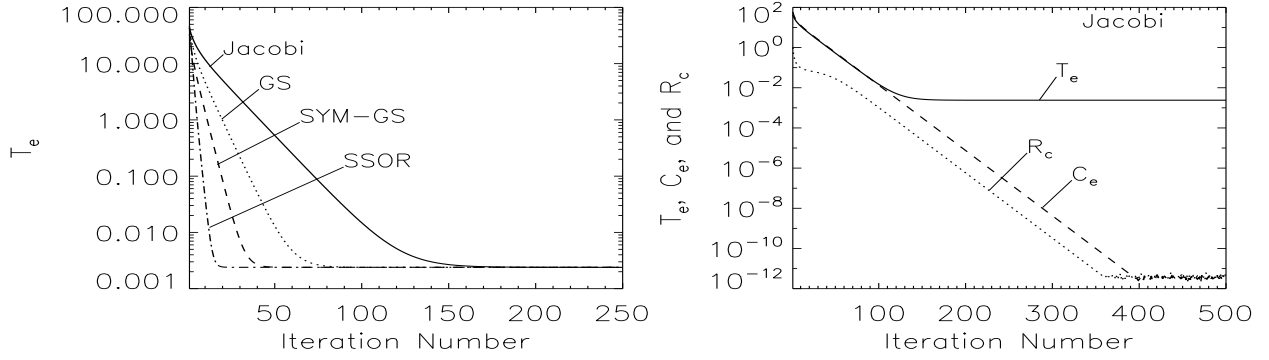}
\caption{Doppler, natural and collisional (type III) redistribution. 
Convergence properties of the various iterative schemes in a 
semi-infinite isothermal 
model atmosphere with $\epsilon=10^{-4}$, $a=10^{-3}$ and a spatial grid 
with 9 points per decade ($\Delta Z=0.25$). The 
different line types are the same as in Fig.~\ref{r1_tercmrc}.
}
\label{r3_tercmrc}
\end{figure*}

Figure~\ref{r3_tercmrc} shows the convergence properties of the different 
iterative schemes, applied here to type III redistribution. Model parameters 
are the same as those for type I redistribution, but now the 
damping parameter $a=10^{-3}$. The angular and depth grids used for 
the computation are exactly the same as those used for type I redistribution. 
However, a non-uniform frequency grid that extends 
up to 1000 Doppler widths from the line center is used (as now $a\ne 0$). 
We note that for type III redistribution 
we obtain $T_e=2.3\times 10^{-3}$, which is nearly the same true error as 
that obtained for the corresponding CRD case (with $a=10^{-3}$). This is 
expected, as it is well known that 
$R_{\rm III,AA}$, in the rest frame of the atom behaves like CRD.\\

\subsection{Doppler and Natural Broadening - type II redistribution}
Physically type II redistribution represents the case of a line 
with an infinitely sharp lower level and an upper level broadened by 
radiative decay only. In the rest frame of the atom the absorption profile 
is a Lorentzian and the scattering is completely coherent. This type of 
scattering problem is essential to model strong resonance lines, formed in 
low density media. 

\begin{figure*}
%\centering  
%\includegraphics[height=5cm,width=16cm]{r2_tecemrc.eps}
\plotone{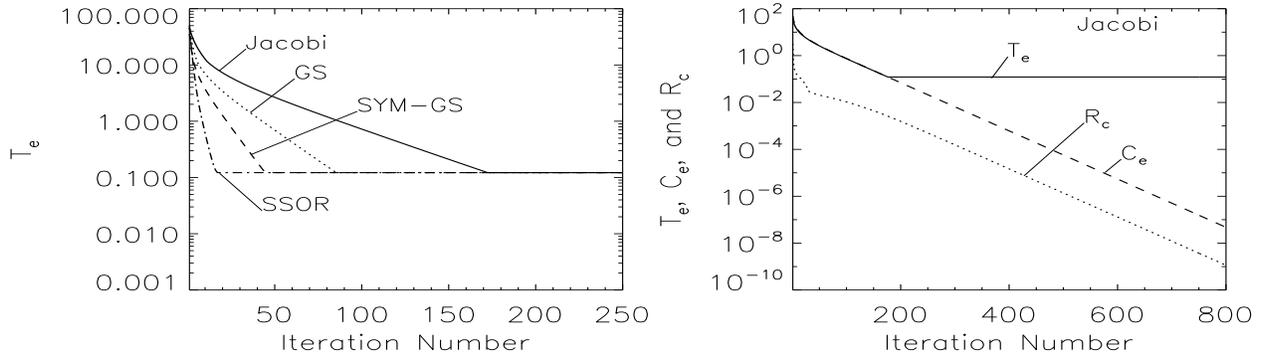}
\caption{Doppler and natural broadening (type II redistribution). 
Convergence properties of the various iterative schemes in a 
semi-infinite isothermal 
model atmosphere with $\epsilon=10^{-4}$, $a=10^{-3}$ and a spatial grid 
with 9 points per decade ($\Delta Z=0.25$). 
The line types and the model parameters are exactly the same as in 
Fig.~\ref{r3_tercmrc}.
}
\label{r2_tercmrc}
\end{figure*}

Figure~\ref{r2_tercmrc} shows the convergence properties of different 
iterative schemes, applied here to the type II redistribution problem. 
The model parameters and the various grids used for the computation are 
the same as those used in \S~4.2 for type III redistribution. We point 
out that the SSOR method works well 
when combined with the FBF technique, while it doesn't work properly 
when combined with the 
CRD-CS method. The reason is probably due to the way the wings are handled 
in the CRD-CS method. For type II redistribution we obtain $T_e=0.12$, 
which is pretty a high value compared to that obtained in the type I, III and 
CRD cases. It is well known that one needs a much more refined frequency grid 
for $R_{\rm II,AA}$ than for the other redistribution functions because the 
asymptotic large scale behavior of the transfer equation for $R_{\rm II,AA}$ 
is like a space and frequency diffusion equation \citep[see][]{hf88}. 
However, we checked that use of a frequency grid 
even finer than the non-uniform frequency grid mentioned above does not change 
the $T_e$ value quoted above. Such a high value of $T_e$ could 
be due to the fact that $R_{\rm II,AA}$ has coherent peaks in the 
wings, while other functions do not have coherent peaks. 
Furthermore, in the case of $R_{\rm II,AA}$ the wings 
cannot be easily thermalized \citep[see][]{hf80}. It is worth to note that 
very far in the wings only diffusion in space remains. Such a regime is 
encountered only in pure $R_{\rm II,AA}$ problems. The presence of a 
background continuum or of some elastic collisions will hide this very 
far wing regime and thereby decreases $T_e$ (see below). 

\begin{figure}
%\centering   
%\includegraphics[height=10cm,width=16cm]{r2_tecemrc_additional.eps}
\plotone{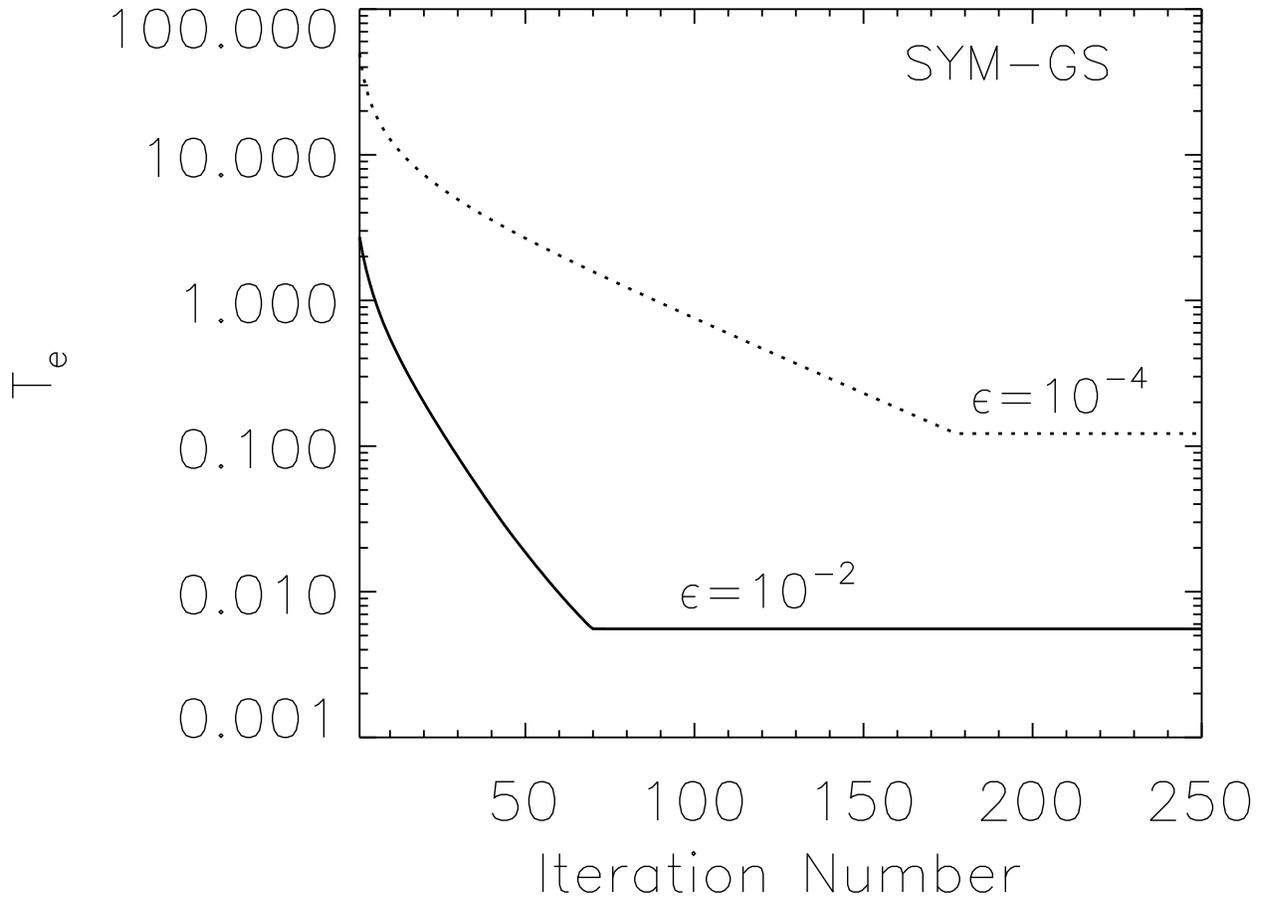}
\caption{Study of the true error for the type II redistribution function 
problem. Different line types\,: solid line ($\epsilon=10^{-2}$), and 
dotted line ($\epsilon=10^{-4}$). 
A semi-infinite isothermal 
model atmosphere with $a=10^{-3}$ and a spatial grid 
with 9 points per decade ($\Delta Z=0.25$) are used. 
}
\label{r2_tercmrc_additional}
\end{figure}

We made a detailed study of the true error for the $R_{\rm II,AA}$ 
redistribution function case using the SYM-GS iterative 
method. Figure~\ref{r2_tercmrc_additional}, shows the true error for 
different  $\epsilon$ values. 
Note that the true error decreases when the non-LTE 
parameter $\epsilon$ increases (i.e., when the 
number of scattering events decreases). 
In Table~\ref{r2_te_depth} we present the true error for different 
resolutions of the depth grid. 
As expected, the true error decreases as the grid resolution increases. 

\begin{figure}
%\centering   
%\includegraphics[height=5cm,width=7.5cm]{r2_truerror_cont.eps}
\plotone{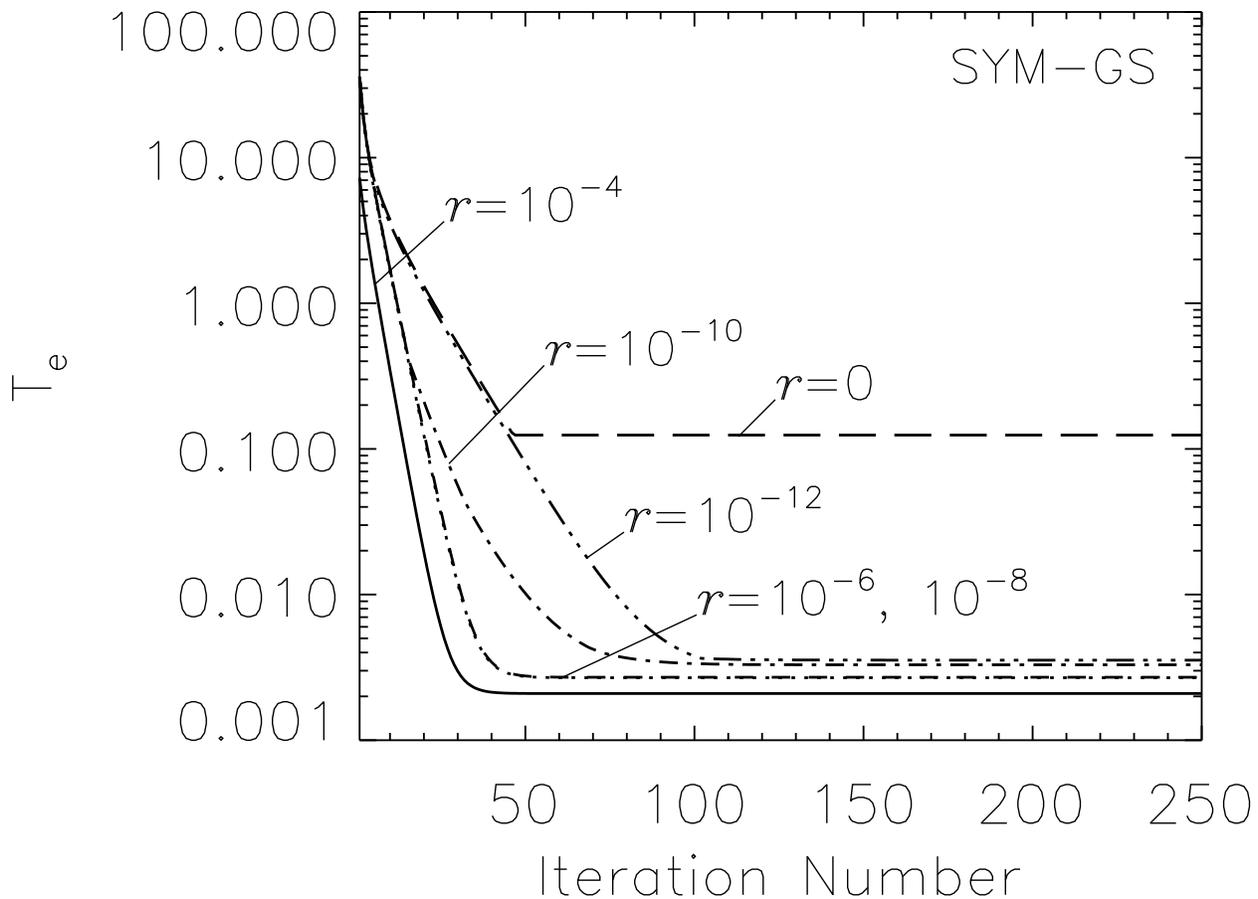}
\caption{Study of the true error for the type II redistribution function case. 
Effect of the continuum parameter $r$. The non-LTE parameter 
$\epsilon=10^{-4}$ and the depth grid spacing $\Delta Z=0.25$. 
Solid line\,: $r=10^{-4}$, dotted line\,: $r=10^{-6}$, 
dashed line\,: $r=10^{-8}$, dot-dashed line\,: $r=10^{-10}$, 
dash-triple-dotted line\,: $r=10^{-12}$, and long-dashed line\,: $r=0$. 
Note that the dotted and dashed lines merge to give a dot-dashed line. 
}
\label{r2_truerror_cont}
\end{figure}

Figure~\ref{r2_truerror_cont} shows the behavior of the true error for type 
II redistribution when a background continuum is included. Clearly, the 
addition of the continuum decreases the true error substantially, as the 
wings can then be thermalized. Note that even with an opacity ratio $r$ 
as small as $10^{-12}$ the true error decreases to nearly 
$3.5\times 10^{-3}$, from $T_e=0.12$ for the pure 
line case. Since in practical problems a background continuum is always 
present, we can conclude that the true error of the numerical methods 
based on operator splitting for type II redistribution can be made 
significantly small. For example, it is $0.2\,\%$ when 
$r=10^{-4}$ and $\Delta Z=0.25$. 

\subsection{Linear combination of $R_{\rm II,AA}$ and $R_{\rm III,AA}$}
We now consider a more realistic case characterized by the following weighted 
combination of type II 
and type III redistribution \citep[e.g.,][]{ste94}\,:
\begin{equation}
R_{\rm AA}(x,x^\prime) = \gamma R_{\rm II,AA}(x,x^\prime) + 
(1-\gamma)R_{\rm III,AA}(x,x^\prime),
\label{r23_mix}
\end{equation}
where $\gamma=1/(1+\Gamma_{\rm E}/\Gamma_{\rm R})$, with $\Gamma_{\rm E}$ 
the elastic collisional rate and $\Gamma_{\rm R}$ the radiative rate. 

\begin{figure}
%\centering  
%\includegraphics[height=5cm,width=7.5cm]{r23_te_gebyr.eps}
\plotone{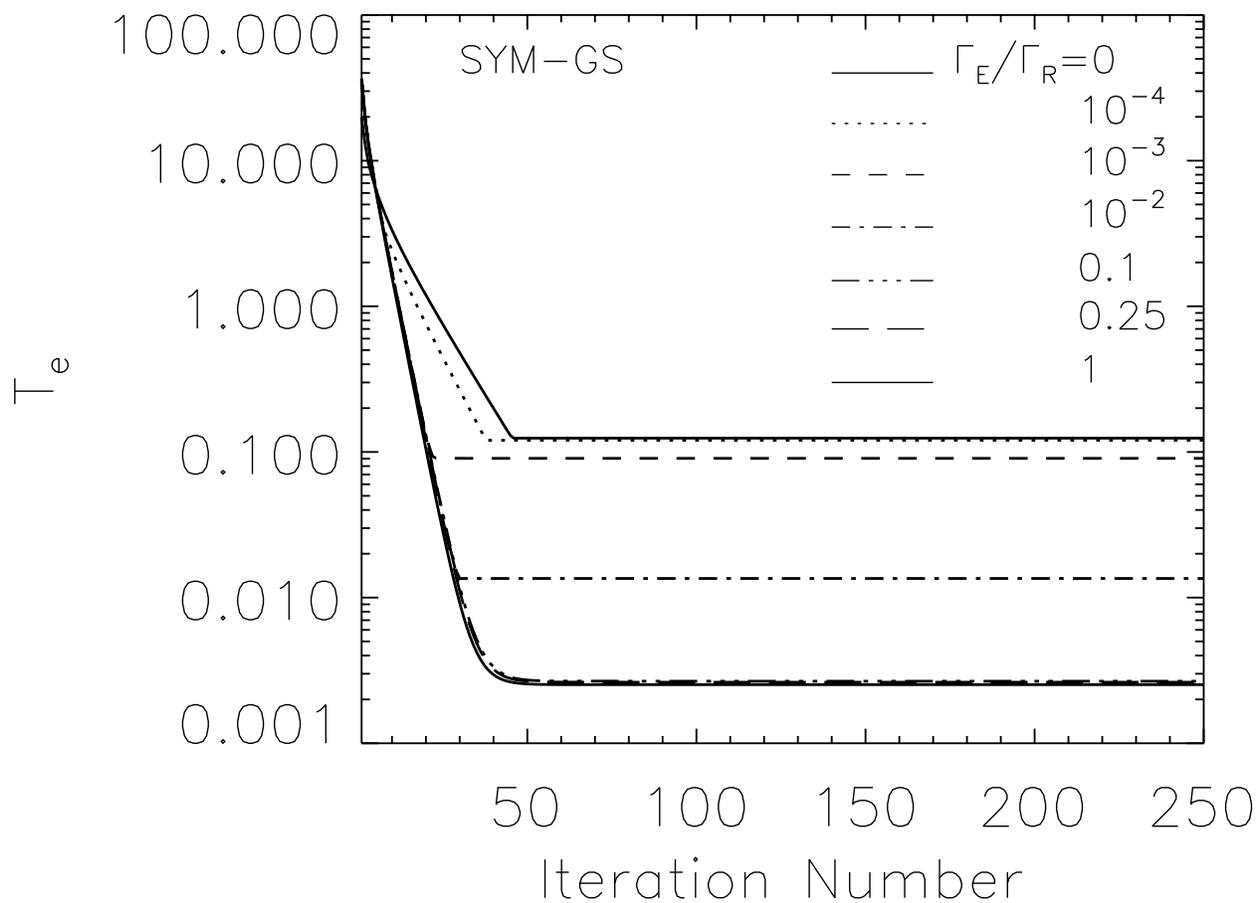}
\caption{Study of the true error for the combined case 
of type II and III redistribution function. 
Effect of the elastic collision parameter $\Gamma_{\rm E}/\Gamma_{\rm R}$. 
The non-LTE parameter $\epsilon=10^{-4}$, the continuum parameter $r=0$ and 
the depth grid spacing 
$\Delta Z=0.25$. The topmost solid line\,: 
$\Gamma_{\rm E}/\Gamma_{\rm R}=0$, dotted line\,:
$\Gamma_{\rm E}/\Gamma_{\rm R}=10^{-4}$, dashed line\,: 
$\Gamma_{\rm E}/\Gamma_{\rm R}=10^{-3}$, 
dot-dashed line\,: $\Gamma_{\rm E}/\Gamma_{\rm R}=10^{-2}$, 
dash-triple-dotted line\,: $\Gamma_{\rm E}/\Gamma_{\rm R}=0.1$, 
long-dashed line\,:
$\Gamma_{\rm E}/\Gamma_{\rm R}=0.25$, and the bottom most solid line\,: 
$\Gamma_{\rm E}/\Gamma_{\rm R}=1$. 
}
\label{r23_te_gebyr}
\end{figure}

Figure~\ref{r23_te_gebyr} shows the behavior of the true error for different 
choices of the elastic collision parameter $\Gamma_{\rm E}/\Gamma_{\rm R}$. The 
true error corresponding to $\Gamma_{\rm E}/\Gamma_{\rm R}=0$ is nothing but 
that corresponding to the pure $R_{\rm II,AA}$ case, which shows the largest 
value for the truncation error. Introducing a small mix of type 
III redistribution through the contribution of elastic collisions results in 
a decrease of the true error. Already for $\Gamma_{\rm E}/\Gamma_{\rm R}=0.1$, 
the true error is nearly the same as that corresponding to the CRD case. This 
again shows that the coherent peaks of $R_{\rm II,AA}$ are responsible for a 
large truncation error in the case of pure type II redistribution without any 
background continuum. 

\section{Polarized PRD radiative transfer equation}
In this paper we restrict ourselves to situations where the radiation field 
is axially symmetric. This condition is satisfied only for one-dimensional 
plane-parallel or spherical atmospheres with either no magnetic field, or 
a micro-turbulent and isotropic field, or a micro-structured magnetic field 
with a fixed inclination and a random azimuth. 
Here we consider the case of a plane-parallel atmosphere with zero magnetic 
field.\footnote{ We remark that a microturbulent magnetic field can 
be taken into account by simply replacing $W_2(J_l,J_u)$ (see \S~5.1. for its 
definition) by $H_2\,W_2(J_l,J_u)$, where $H_2$ is the so-called Hanle 
depolarization factor. $H_2$ is unity when the magnetic strength is zero. The 
explicit form of $H_2$ for an isotropic magnetic field and a horizontal 
magnetic field with random azimuth can be found in \citet{ste94} and 
\citet{ll04}.} 
An axially symmetric polarized 
radiation field is described by the Stokes parameters $I$ and $Q$ 
\citep[see][]{chandra50}, where $I$ denotes the intensity and $Q$ the 
linear polarization (i.e., the difference between the intensity components 
parallel and perpendicular to a given reference direction in the plane perpendicular to 
the direction of the ray under consideration). In this paper, the positive 
$Q$ direction is defined in the plane containing the direction of the ray and the vertical Z-axis. 
The one-dimensional transfer equation for the Stokes vector components 
${I}_{x\mu,j}=(I,Q)$ for $j=0,1$ is given by 
\begin{equation}
{{\rm d}  \over {\rm d} \tau} {I}_{x\mu,j}(\tau)=
{I}_{x\mu,j}(\tau) - {S}_{x\mu,j}(\tau).
\label{prte-comp}
\end{equation}
The source vector components ${S}_{x\mu,j}=({S}^{I}_{x\mu},{S}^{Q}_{x\mu})$ 
for $j=0,1$ are of the form 
\begin{equation}
{S}_{x\mu,j} = {\phi_x {S}_{lx\mu,j} + r B{U}_j\over \phi_x+r} ,
\label{vectorsource_total_comp}
\end{equation}
where ${U}_j=(1,0)$ for $j=0,1$, and the line source vector components 
${S}_{lx\mu,j}$ are given by \citep[e.g.,][]{reeandsal82}
\begin{eqnarray}
{S}_{lx\mu,j}&=&\epsilon B {U}_j
+  \int_{-\infty}^{+\infty} {\rm d}x^\prime {1\over 2} 
\int_{-1}^{+1} {\rm d}\mu^\prime \nonumber \\ &&\!\!\!\!\!\!\!\!\!\!\times
\sum_{j^\prime=0}^{1}\left[{\bf R}(x,x^\prime; \mu, \mu^\prime) \right]_{jj^\prime}
{I}_{x^\prime\mu^\prime,j^\prime}.
\label{linesourcevector}
\end{eqnarray}
In the above equation 
$\left[{\bf R}(x,x^\prime; \mu, \mu^\prime)\right]_{jj^\prime}$ are 
the elements of the scattering redistribution 
matrix ${\bf R}(x,x^\prime; \mu, \mu^\prime)$ for the non-magnetic case 
\citep{reeandsal82,dh88}. 
In the following subsections we first discuss the redistribution matrix 
for the non-magnetic case, and then present the decomposition technique 
proposed by \citet{hf07}. This is because the iterative algorithms given 
in this paper are based on the ensuing equations deduced in \S~5.2. 

\subsection{Redistribution matrix}
A hybrid approximation to ${\bf R}(x,x^\prime; \mu, \mu^\prime)$ was 
introduced by \citet{reeandsal82}\,:
\begin{equation}
{\bf R}(x,x^\prime; \mu, \mu^\prime) = (1-\epsilon) g^k_{xx^\prime} 
{\bf P}(\mu,\mu^\prime),
\label{hybrid_approx}
\end{equation}
where the phase matrix ${\bf P}(\mu,\mu^\prime)$ is given by 
\citep[e.g.,][]{ll04,bom97}
\begin{equation}
{\bf P}(\mu,\mu^\prime) = \sum_{K=0,2} W_{K}(J_l,J_u) {\bf P}^{K}_{\rm R}
(\mu,\mu^\prime).
\label{gen_phase_mat}
\end{equation}
The coefficient $W_{0}(J_l,J_u)=1$, with $J_l$ and $J_u$ being the total 
angular momentum quantum numbers of the lower and upper levels, respectively. 
The coefficient $W_{2}(J_l,J_u)$ characterizes the maximum linear polarization 
that can be produced in the line. In the case of a normal Zeeman triplet 
($J_l=0, J_u=1$), $W_{2}(J_l,J_u)=1$, and ${\bf P}(\mu,\mu^\prime)=
{\bf P}_{\rm R}(\mu,\mu^\prime)$ is the so-called Rayleigh phase matrix. 
Even though the figures of this paper correspond to the case of 
a normal Zeeman triplet, we present the equations for the more general 
case of arbitrary values of $W_{K}(J_l,J_u)$. The Rayleigh phase matrix 
multipolar components ${\bf P}^{K}_{\rm R}(\mu,\mu^\prime)$ are given by 
\citep[see][written here for the azimuthally symmetric case]{lan84}
\begin{equation}
\left[{\bf P}^{K}_{\rm R}(\mu,\mu^\prime)\right]_{jj^\prime}=
\tilde{\mathcal T}^K_0(j,\theta)\tilde{\mathcal T}^K_0(j^\prime,\theta^\prime),
\label{pkr}
\end{equation}
where $j,j^\prime=0,1$. The notation $\tilde{\mathcal T}^K_Q(j,\theta)$ was 
introduced by \citet[][see her Equation~(28)]{hf07}, where for each $K$, $Q$ 
takes values between $-K$ to $+K$ in steps of unity. These quantities are 
related to the irreducible tensors for polarimetry 
${\mathcal T}^K_Q(j,\bm{\Omega})$ 
introduced by \citet{lan84}, where $\bm{\Omega}=(\theta,\chi)$ denote the 
ray direction \citep[see][]{hf07}. 
Since here we are dealing with the azimuthally symmetric case, the relevant 
quantities corresponding to $Q=0$ are \citep[see Table~5.6 of][]{ll04}
\begin{eqnarray}
\tilde{\mathcal T}^0_0(0,\theta)=1\,; \quad 
\tilde{\mathcal T}^2_0(0,\theta)={1\over 2\sqrt{2}} (3\mu^2-1),\nonumber \\
\tilde{\mathcal T}^0_0(1,\theta)=0\,; \quad 
\tilde{\mathcal T}^2_0(1,\theta)=-{3\over 2\sqrt{2}} (1-\mu^2).
\label{tk0_expression}
\end{eqnarray}

In Equation~(\ref{hybrid_approx}), $\epsilon=\Gamma_{\rm I}/
(\Gamma_{\rm I}+\Gamma_{\rm R})$ with $\Gamma_{\rm I}$ being the inelastic 
collisional rate and $\Gamma_{\rm R}$ the radiative rate. However, 
Equation~(\ref{hybrid_approx}) is only an approximate form of the 
redistribution matrix as it does not take into account the effect of elastic 
($\Gamma_{\rm E}$) and depolarizing ($D^{(K)}$) collisional rates. The first 
quantum mechanical calculation of the redistribution matrix for the resonance 
polarization, taking into account the effect of elastic collisions was 
performed by \citet{osc72}. Starting from the work of \citet{osc72}, 
\citet{dh88} derived a tractable analytic expression of the redistribution 
matrix. It is worth to note that the redistribution
matrix that was derived by \citet{dh88}, is very general, namely, it depends
on the angle-dependent redistribution functions of \citet{hum62}. However
for computational simplicity, following \citet{reeandsal82},
\citet[][see also Faurobert-Scholl 1992]{knn94} used the
angle-averaged version of the Domke-Hubeny (DH) redistribution matrix. 
Following \citet{bom97} we write this 
redistribution matrix as follows\,:
\begin{eqnarray}
&{\bf R}_{\rm DH}(x,x^\prime; \mu,\mu^\prime)=  
\sum_{K=0,2} W_{K}(J_l,J_u) \nonumber \\ 
&\times \left\{\alpha\, g^{\rm II}_{xx^\prime} + 
[\beta^{(K)}-\alpha]\, g^{\rm III}_{xx^\prime}\right\}
{\bf P}^{K}_{\rm R}(\mu,\mu^\prime),
\label{dh_redist}
\end{eqnarray}
where the branching ratios $\alpha$ and $\beta^{(K)}$ are given by 
\begin{equation}
\label{branch_alpha}
\alpha={\Gamma_{\rm R}\over \Gamma_{\rm R}+\Gamma_{\rm I}+\Gamma_{\rm E}},
\end{equation}
\begin{equation}
\beta^{(K)}={\Gamma_{\rm R}\over \Gamma_{\rm R}+\Gamma_{\rm I}+D^{(K)}}.
\label{branch_beta}
\end{equation}
Note that $D^{(0)}=0$, and also that the factor $(1-\epsilon)$ is contained 
in the branching ratios. 

It is worth to clarify certain important points related to these branching 
ratios \citep[see also][]{knn94}. In astrophysics one expects that the 
branching ratios add up to unity. However, from Equation~(\ref{dh_redist}) 
we see that the branching ratios 
add up to give $\left[\alpha+\beta^{(K)}-\alpha\right]=\beta^{(K)}$, which 
for $K=0$ is nothing but $(1-\epsilon)$ and for $K=2$ is 
$(1-\epsilon)/\left[1+\delta^{(2)}(1-\epsilon)\right]$ with 
$\delta^{(2)}=D^{(2)}/\Gamma_{\rm R}$. We note that these are indeed the 
factors that appear in the line source function expressions for Stokes $I$ 
and Stokes $Q$ (namely $S^0_0$ and $S^2_0$ or $\rho^0_0$ and $\rho^2_0$) 
in the CRD case formulated \citep[see][]{jtbrms99,ll04}. It is important to 
note that some authors \citep[e.g.,][]{fau92,knn94}, write the factor 
$(1-\epsilon)$ 
before the second term of Equation~(\ref{linesourcevector}) and renormalize the 
branching ratios $\alpha$ and $\beta^{(K)}$ by $(1-\epsilon)$, namely
\begin{eqnarray}
(\alpha)^{\rm old}={\alpha \over 1-\epsilon}=
{\Gamma_{\rm R}+\Gamma_{\rm I} \over 
\Gamma_{\rm R}+\Gamma_{\rm I}+\Gamma_{\rm E}}, \nonumber \\
\left[\beta^{(K)}\right]^{\rm old}={\beta^{(K)} \over 1-\epsilon}=
{\Gamma_{\rm R}+\Gamma_{\rm I} \over 
\Gamma_{\rm R}+\Gamma_{\rm I}+D^{(K)}}.
\end{eqnarray}

The approximate form of ${\bf R}(x,x^\prime; \mu,\mu^\prime)$ given in 
Equation~(\ref{hybrid_approx}) was used in plane-parallel polarized radiative 
transfer by \citet{reeandsal82} and \citet{fau87,fau88}, with 
$g^{\rm II}_{xx^\prime}$. These authors used Feautrier's (1964) method 
to solve the polarized transfer equation. 
A discrete space method was used by \citet{knn86,knn88,knn89} for the same 
problem but in spherical atmospheres. \citet{mck84} used an integral equation 
approach to solve the same problem with a linear combination of 
$g^{\rm II}_{xx^\prime}$ and $g^{\rm III}_{xx^\prime}$ 
(see Equation~(\ref{r23_mix})). The DH redistribution matrix given in 
Equation~(\ref{dh_redist}) was used in plane-parallel polarized radiative transfer 
computations by \citet{fau92,fau93}. The same problem was 
solved by \citet{knn94,knn95} but in spherical atmospheres. As already 
mentioned in the introduction such methods are computationally expensive. 

A Jacobi based ALI method to solve the polarized radiative transfer equation 
with the hybrid approximation for the redistribution matrix and 
$g^{\rm II}_{xx^\prime}$ was developed by \citet{fpmf97}. They extended 
the CRD-CS method of PA95 to scattering polarization. \citet{jtbrms99} 
generalized the symmetric GS and SOR methods of TF95 to CRD polarized 
radiative transfer, with the relevant equations formulated within the 
framework of the quantum theory of spectral line polarization described in the monograph 
by \citet{ll04}. In this paper we generalize these symmetric GS and 
SOR methods to solve the above-mentioned PRD problem. We consider 
both, the hybrid approximation and the DH redistribution matrix. In 
the next section we present the Jacobi, GS and SOR iterative methods to 
solve polarized radiative transfer problems with the DH redistribution matrix. 

\subsection{Decomposition in the irreducible basis}
From Equations~(\ref{vectorsource_total_comp}) and (\ref{linesourcevector}) we see 
that unlike the unpolarized case, the line source vector components now depend 
not only on the frequency $x$ but also on the orientation $\mu$ of the 
radiation beams. In the case of CRD the line source vector components depend 
only on $\mu$ and are independent of $x$. To reduce the computational cost, 
\citet[][see also Paletou \& Faurobert-Scholl 1997]{fauetal97} used a 
factorized form of ${\bf P}(\mu,\mu^\prime)$ given by \citet{iva95}, which 
allowed them to transform or reduce the polarized CRD transfer equation to 
a $2\times 2$ basis wherein the source vector components 
are independent of $\mu$. To this $2\times 2$ matrix transfer equation 
they applied a Jacobi iterative scheme to solve the problem. 

The factorization of the Rayleigh phase matrix into a 
product of two $2\times 2$ matrices that depend separately on $\mu$ and 
$\mu^\prime$ is not unique \citep[see][]{hf07}. 
Such a factorization comes out naturally if one uses the ${\mathcal T}^K_Q(j,
{\bm {\Omega}})$ irreducible tensors \citep[see][]{lan84} to derive the 
Rayleigh phase matrix (see Equation~(\ref{pkr})). Using the irreducible tensors 
${\mathcal T}^K_Q(j,{\bm \Omega})$, \citet{hf07} provided a simple way of 
transforming or reducing the Stokes vector components to irreducible tensors 
in the case of the Hanle effect regime. Such a transformation is referred 
to as the ``decomposition'' of the Stokes vector components. We note that 
such a decomposition comes out naturally in the density matrix theory of 
spectral line polarization \citep[see][]{ll04}. Furthermore, it is well 
known that the density matrix and the 
scattering formalisms (that we adopt in this paper) are equivalent for a 
two-level atom without lower-level polarization and stimulated emission. 
However, in this paper we use the decomposition technique proposed by 
\citet{hf07}, but applied here to the axially symmetric case. For clarity, we 
present some important steps of this 
decomposition. For more details the reader is referred to \citet{hf07}. 

For the azimuthally symmetric case the Stokes vector component 
decomposition given by \citet{hf07} takes the 
following form (in the notations used in this paper)\,: 
\begin{equation}
I_{x\mu,j}=\sum_{K=0,2}\tilde{\mathcal T}^K_0(j,\theta)\,
(I_{x\mu})^K_0,
\label{ixmujtoixmuk0}
\end{equation}
with similar equations relating $U_j$, $S_{x\mu,j}$ and $S_{lx\mu,j}$ to 
$U^K_0$, $(S_{x})^K_0$ and $(S_{lx})^K_0$, respectively. 
Note that $U^0_0=1$ and $U^2_0=0$. The quantities $(I_{x\mu})^K_0$ 
and $(S_{x})^K_0$ are called the irreducible tensor components of the Stokes 
and the source vector components, respectively. 

Substituting Equations~(\ref{pkr}), (\ref{dh_redist}) and 
(\ref{ixmujtoixmuk0}) in 
Equations~(\ref{prte-comp})--(\ref{linesourcevector}), 
it can be shown that 
$(I_{x\mu})^K_0$ satisfies a transfer equation similar to 
Equation~(\ref{prte-comp}) but with $I_{x\mu,j}$ and $S_{x\mu,j}$ replaced 
by $(I_{x\mu})^K_0$ and $(S_{x\mu})^K_0$, respectively. Furthermore, 
$(S_{x\mu})^K_0$ is given by Equation~(\ref{vectorsource_total_comp}) but 
with $S_{lx\mu,j}$ and $U_j$ replaced by $(S_{x\mu})^K_0$ and $U^K_0$, 
respectively. The irreducible components of the line source vector are 
now given by 
\begin{equation}
({S}_{lx})^K_0=\epsilon B {U}^K_0 
+  W_K(J_l,J_u) (\bar J_x)^K_0, 
\label{slxk0}
\end{equation}
where
\begin{eqnarray}
(\bar J_x)^K_0&=&
\int_{-\infty}^{+\infty} {\rm d}x^\prime \nonumber \\ &&
\!\!\!\!\!\!\!\!\!\!\!\!\!\!\!\!\!\!\!\!\!\!\!\!\!\!\!\!\!\!\!\!\!\!\!\!
\times \left\{\alpha g^{\rm II}_{xx^\prime}
+\left[\beta^{(K)}-\alpha\right] g^{\rm III}_{xx^\prime}\right\} 
(J_{x^\prime})^K_0. 
\label{linesourcevector_comp}
\end{eqnarray}
In the above equation the angle integrated irreducible tensor is given by 
\begin{equation}
(J_{x^\prime})^K_0=\sum_{K^\prime=0,2} {1\over 2} \int_{-1}^{+1}
\Psi^{KK^\prime}_0(\mu^\prime)
({I}_{x^\prime\mu^\prime})^{K^\prime}_0(\tau){\rm d}\mu^\prime,
\label{jk0-ik0}
\end{equation}
where
\begin{equation}
\Psi^{KK^\prime}_0(\mu^\prime)=\sum_{j^\prime=0}^{1}
\tilde{\mathcal T}^K_0(j^\prime,\theta^\prime)
\tilde{\mathcal T}^{K^\prime}_0(j^\prime,\theta^\prime).
\label{psikkprime0}
\end{equation}
Clearly, the advantage of this decomposition is that the irreducible 
tensor components of the source vectors are now independent of the orientation 
$\mu$ of the radiation beam. 

Following \citet{hf07} we now introduce the 2-component Stokes 
and source vectors in the irreducible basis
\begin{equation}
{\bm{\mathcal I}}_{x\mu} = \left[(I_{x\mu})^0_0,\, (I_{x\mu})^2_0\right]^{\rm T};
\ \  
{\bm{\mathcal S}}_{x} = \left[(S_{x})^0_0,\, (S_{x})^2_0\right]^{\rm T}.
\label{2-vectors}
\end{equation}
In the above vector notation, the transfer equation in the irreducible basis 
can be written as 
\begin{equation}
{{\rm d}  \over {\rm d} \tau} {\bm{\mathcal I}}_{x\mu}(\tau)=
{\bm{\mathcal I}}_{x\mu}(\tau) - {\bm{\mathcal S}}_{x}(\tau),
\label{prte-vectori}
\end{equation}
where ${\bm{\mathcal S}}_{x}$ is given by Equation~(\ref{vectorsource_total_comp}), 
but with $S_{lx\mu,j}$ and $U_j$ replaced by ${\bm{\mathcal S}}_{lx}$ and 
${\bm{\mathcal U}}$, respectively. Here ${\bm{\mathcal U}}=(1,0)^{\rm T}$, 
and  
\begin{equation}
{\bm{\mathcal S}}_{lx}=\epsilon B {\bm{\mathcal U}} 
+  {\bf W} \ \overline{{\bm{\mathcal J}}}_x. 
\label{vectorsl}
\end{equation}
In the above equation 
\begin{equation}
\overline{{\bm{\mathcal J}}}_x = \int_{-\infty}^{+\infty} {\bf N}_{xx^\prime} 
{\bm{\mathcal J}}_{x^\prime}\,{\rm d}x^\prime,
\label{jbar_vector}
\end{equation}
where
\begin{equation}
{\bf N}_{xx^\prime} = g^{\rm II}_{xx^\prime} \alpha {\bf E} 
+ g^{\rm III}_{xx^\prime} \left( {\bf{\mathcal B}} - \alpha {\bf E}\right)
,
\label{nxxprime}
\end{equation}
and the 2-component mean intensity vector 
\begin{equation}
{\bm{\mathcal J}}_{x^\prime}={1\over 2}\int_{-1}^{+1} 
{\bm \Psi}(\mu^\prime) \,{\bm{\mathcal I}}_{x^\prime\mu^\prime} \,
{\rm d}\mu^\prime.
\label{vectorj}
\end{equation}
In Equations~(\ref{nxxprime}) and (\ref{vectorj}), 
${\bf E}$ is a $2\times 2$ identity matrix, while 
${\bf W}={\rm diag}[W_0,W_2]$ and 
${\bm{\mathcal B}}={\rm diag}[\beta^{(0)},\beta^{(2)}]$ are $2 \times 2$ 
matrices. Note that since the matrix ${\bm{\mathcal B}}$ is diagonal, 
the matrix ${\bf N}_{xx^\prime}$ is also diagonal. The explicit form of the 
$2\times 2$ matrix ${\bm \Psi}$ formed by the elements 
$\Psi^{KK^\prime}_0$ can be found in Appendix A of \citet{hf07}. 
In the next section we apply the Jacobi, GS and SOR iterative schemes 
to Equations~(\ref{prte-vectori})--(\ref{vectorj}). 

\section{Iterative methods for polarized PRD radiative transfer}
The formal solution of Equation~(\ref{prte-vectori}) is given by 
Equation~(\ref{formal_solution}), but with $I_{x\mu}$, $\Lambda_{x\mu}$, 
$S_{x}$ and $T_{x\mu}$ replaced by 
${\bm{\mathcal I}}_{x\mu}$, ${\bf \Lambda}_{x\mu}$, 
${\bm{\mathcal S}}_{x}$, and ${\bm T}_{x\mu}$, respectively. 
Here ${\bm T}_{x\mu}$ is the transmitted 2-component Stokes vector due to 
the incident radiation at the boundaries, and ${\bf \Lambda}_{x\mu}$ is a 
$2N\times 2N$ operator. For given depth indices $i,i^\prime$, 
${\bf \Lambda}_{x\mu,ii^\prime}$ is a $2\times 2$ block. We use again the 
short-characteristics method as the  formal solver, but now applied 
to the vector transfer equation~(\ref{prte-vectori}). 

As in Equation~(\ref{scat_int_iter}), we now write the 2-component mean 
intensity vector as 
\begin{eqnarray}
{\bm{\mathcal J}}_{x,i}&=&{\bf \Lambda}_{x,i1}{\bm{\mathcal S}}^{a}_{x,1} 
+\cdots +{\bf \Lambda}_{x,ii-1}{\bm{\mathcal S}}^{a}_{x,i-1} \nonumber\\
&&+{\bf \Lambda}_{x,ii}{\bm{\mathcal S}}^{b}_{x,i} 
+{\bf \Lambda}_{x,ii+1}{\bm{\mathcal S}}^{c}_{x,i+1} + 
\cdots \nonumber \\
&&+{\bf \Lambda}_{x,iN}{\bm{\mathcal S}}^{c}_{x,N} + {\bm T}_{x,i}\ ,
\label{2-comp_mean_int_iter}
\end{eqnarray}
where ${\bm T}_{x,i}$ is given by Equation~(\ref{vectorj}) but with ${\bm{\mathcal 
I}}_{x\mu}$ replaced by ${\bm T}_{x\mu}$. The $2\times 2$ matrix 
${\bf \Lambda}_{x,ii^\prime}$ is given by 
\begin{equation}
{\bf \Lambda}_{x,ii^\prime} = {1\over 2} \int_{-1}^{+1} {\bm \Psi}(\mu)\,
{\bf \Lambda}_{x\mu,ii^\prime}\,{\rm d}\mu.
\label{vector_lambda}
\end{equation}
Note that unlike the unpolarized case (see Equation~(\ref{scat_int_iter})), at each 
depth we now have to perform a matrix operation involving a $2\times 2$ 
matrix and a $2$ column vector. In the following subsections we successively 
present the Jacobi, GS and SOR iterative schemes. 

\subsection{Jacobi Iterative scheme}
It is straightforward to generalize the Jacobi scheme discussed in \S~3.1 
to the scattering polarization case. With this iterative scheme, 
the equations for the 2-component line source vector corrections are 
given by 
\begin{equation}
\delta {\bm{\mathcal S}}_{lx,i} - {\bf W}\int_{-\infty}^{+\infty} 
{\bf N}_{xx^\prime} 
p_{x^\prime} {\bf \Lambda}_{x^\prime,ii}\, 
\delta {\bm{\mathcal S}}_{lx^\prime,i}\,
{\rm d}x^\prime = {\bm{\mathcal R}}_{x,i}\,,
\label{vectordeltaslx_correction}
\end{equation}
where
\begin{equation}
{\bm{\mathcal R}}_{x,i} = \epsilon \,B\, {\bm{\mathcal U}} + {\bf W}\ 
\overline{{\bm{\mathcal J}}}^{\rm \,old}_{x,i} 
-{\bm{\mathcal S}}^{\rm old}_{lx,i}\,.
\label{vectorrxi}
\end{equation}
As discussed in \S~3.1. the system of linear 
equation~(\ref{vectordeltaslx_correction}) can be solved by the FBF method, 
namely
\begin{equation}
{\bm{\mathcal A}}\, \delta {\bm{\mathcal S}} = {\bm{\mathcal R}},
\label{vector_fbf}
\end{equation}
where at each depth point $i$, ${\bm{\mathcal R}}$ is a vector of length 
$2N_x$, and the matrix ${\bm{\mathcal A}}$ is of dimension $2N_x\times 2N_x$. 
For a given depth point $i$,  and given frequencies $x,x^\prime$, 
${\bm{\mathcal A}}$ is a $2\times 2$ block denoted by ${\bm{\mathcal A}}^2$, 
and given by the expression
\begin{equation}
{\bm{\mathcal A}}^2 = \delta_{mn}\, {\bf E} - {\bf W}\,{\bf N}_{mn}\, p_n \,
{\bf \Lambda}_{n,ii}\,; \ \ m,n=1,\cdots, N_x.
\label{vector_A_mat}
\end{equation}
Clearly, the above FBF method is numerically much more expensive compared 
to the unpolarized case, as the size of the matrix ${\bm{\mathcal A}}$ 
is now twice larger. The FBF method was actually generalized by 
\citet{sametal08} for the weak field regime of the Hanle effect (with PRD). 
This method has also been used by \citet{hfetal09} for the Hanle 
effect of a turbulent magnetic fields with a finite correlation length 
(and with CRD). 

As already discussed in \S~3.1. for the unpolarized case, the CRD-CS 
method is computationally less expensive, but is as robust as the FBF method. 
This CRD-CS method was extended to scattering polarization (non-magnetic) by 
\citet{fpmf97} for the hybrid approximation that uses 
$g^{\rm II}_{xx^\prime}$. This method was later extended to the Hanle effect 
case by \citet{fluetal03}. They used the weak field Hanle redistribution 
matrix of \citet{bom97}, the so-called approximation-III, which reduces to 
the DH redistribution matrix for the zero magnetic field case. In the 
following subsection we briefly recall the CRD-CS method of \citet{fluetal03}, 
applied here to non-magnetic case. The main difference with the equations 
presented in this paper is that 
we use the decomposition technique of \citet{hf07}, while \citet{fluetal03} 
use the traditional Fourier-azimuthal expansion technique discussed in 
\citet{knnetal98}. 

\subsubsection{CRD-CS or core-wing method for the DH redistribution matrix}
As discussed in \S~3.1.1 and 3.1.2, in 
Equation~(\ref{vectordeltaslx_correction}) we approximate 
$g^{\rm II}_{xx^\prime}$ by 
Equation~(\ref{crdcs}) and $g^{\rm III}_{xx^\prime}$ by $\phi_{x^\prime}$ in 
the line core ($x\le x_c$), but set it to zero in the wings ($x > x_c$). 
This gives (using Equation~(\ref{nxxprime}))
\begin{equation}
\delta {\bm{\mathcal S}}_{lx,i}={ {\bm{\mathcal R}}_{x,i} + (1-\alpha_x) \,
{\bf W}\, {\bm{\mathcal B}}\, \Delta {\bm{\mathcal T}}_i \over 
1 - \alpha_x\, \alpha\, p_x\, {\bf W}\, {\bf \Lambda}_{x,ii}}\,.
\label{vector_line_source_corrections_crdcs}
\end{equation}
Following 
\citet[][see also Paletou \& Faurobert-Scholl 1997; Nagendra et al. 1999]{fluetal03}, one can show that
\begin{equation}
\Delta {\bm{\mathcal T}}_i = \left[{\bf E} - \int_{-x_c}^{+x_c} {\rm d}x
\,\phi_x\, p_x\, {\bf \Lambda}_{x,ii}\, \cdot\, {\bf W}\, {\bm{\mathcal B}}
\right]^{-1}\, \bar{\bm{\mathcal R}}_i,
\label{vector_deltat}
\end{equation}
where
\begin{equation}
\bar {\bm{\mathcal R}}_i = \int_{-x_c}^{+x_c} \phi_x\, p_x\, 
{\bf \Lambda}_{x,ii} \, {\bm{\mathcal R}}_{x,i}\,
{\rm d}x.
\label{vector_rbar}
\end{equation}
Note from Equation~(\ref{vector_line_source_corrections_crdcs}) that in 
the core (when $\alpha_x=0$) the denominator reduces to unity 
(i.e., we are left with a simple summation), while in the wings the 
term multiplying $(1-\alpha_x)$ appears as a frequency independent quantity. 

\subsection{GS and SOR iterative schemes}
These radiative transfer methods were developed by 
\citet{jtbrms99} for the non-magnetic and micro-turbulent field CRD cases and 
by \citet{rmsjtb99} for the CRD case of a deterministic 
magnetic field in the Hanle effect regime. Here we present the 
generalization of these methods to the PRD problem of resonance line 
polarization in the absence or in the presence of a weak magnetic field 
which does not break the axial symmetry of the problem (e.g., a microturbulent 
field). 

The GS iterative scheme is obtained by choosing $c={\rm old}$ and 
$a=b={\rm new}$ in Equation~(\ref{2-comp_mean_int_iter}), which gives 
\begin{equation}
{\bm{\mathcal J}}_{x,i}={\bm{\mathcal J}}^{\rm old+new}_{x,i} + 
{\bf \Lambda}_{x,ii}\,\delta {\bm{\mathcal S}}_{x,i}\,,
\label{vectorj_gs}
\end{equation}
where ${\bm{\mathcal J}}^{\rm old+new}_{x,i}$ is the 2-component mean 
intensity vector calculated using `new' values of the source vector 
${\bm{\mathcal S}}_{x,i}$ at grid points $1,2,\cdots,i-1$, and the 
`old' values at $i,i+1,\cdots,N$. The line source vector corrections 
are given by Equation~(\ref{vectordeltaslx_correction}), but with 
\begin{equation}
{\bm{\mathcal R}}_{x,i} = \epsilon \,B\, {\bm{\mathcal U}} + {\bf W}\ 
\overline{{\bm{\mathcal J}}}^{\rm \,old+new}_{x,i}
-{\bm{\mathcal S}}^{\rm old}_{lx,i}\,.
\label{vectorrxi_gs}
\end{equation}
The line source vector corrections can be computed using either the FBF or the 
CRD-CS methods discussed in \S~6.1. Note that the GS as well as the SYM-GS 
iterative algorithms discussed in \S~3.2 can be extended 
straightforwardly to the polarized case. The only difference is that we are now 
dealing with a 2-component source vector, with intensity as well as mean 
intensity vectors, and with an approximate lambda operator which is now a 
$2\times 2$ block for any given depth and frequency. For example, in the 
several correction terms that one needs to consider in a SYM-GS algorithm, 
namely Equations~(\ref{gs_correction_scat_int})--(\ref{gssym_correction_int}), 
we have to replace simply the unpolarized quantities 
$J_{x,i}$, $\bar J_{x,i}$ and $S_{x,i}$ by the polarized 
2-component vectors ${\bm{\mathcal J}}_{x,i}$, 
$\overline{{\bm{\mathcal J}}}_{x,i}$ and 
${\bm{\mathcal S}}_{x,i}$, respectively, to be able to apply those equations 
to the polarized case. Therefore, unlike in the unpolarized case we now have 
to do several matrix manipulations (see, e.g., Equations~(\ref{vector_fbf}) and 
(\ref{vector_line_source_corrections_crdcs})). 

The SOR iterative scheme is obtained by doing the corrections as follows\,:
\begin{equation}
\delta {\bm{\mathcal S}}^{\rm SOR}_{lx,i} = \omega\,\delta 
{\bm{\mathcal S}}^{\rm GS}_{lx,i}.
\label{vector_sor}
\end{equation}

\begin{figure*}
%\centering   
%\includegraphics[height=7.5cm,width=16cm]{lambda_element.eps}
\plotone{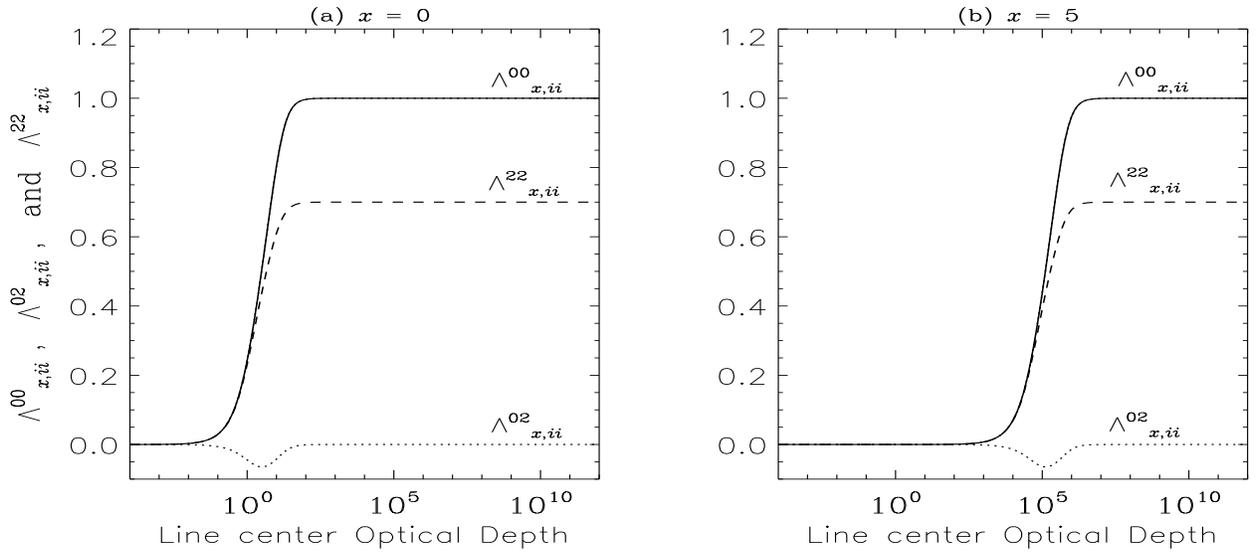}
\caption{Variation of the diagonal elements of the monochromatic lambda 
operator with the line center optical depth. A semi-infinite model atmosphere 
with no continuum ($r=0$) and damping parameter $a=10^{-3}$ are used. 
The solid line corresponds to $\Lambda^{00}_{x,ii}$, the dotted line to 
$\Lambda^{02}_{x,ii}$ and the dashed line to $\Lambda^{22}_{x,ii}$. 
}
\label{lambda_element}
\end{figure*}

As already noted, all the three iterative schemes (Jacobi, GS and SOR)
involve matrix operations for the computation of the line source vector 
corrections (see Equations~(\ref{vector_fbf}) and 
(\ref{vector_line_source_corrections_crdcs})). A smart strategy to avoid such 
matrix computations, and thereby speed up the iterative methods, was given 
by \citet{jtbrms99}. To 
describe this strategy in some detail, we now write the $2\times 2$ 
approximate operator ${\bf \Lambda}_{x,ii}$ as follows\,:
\begin{eqnarray}
{\bf \Lambda}_{x,ii}=\left(
\begin{array}{cc}
\Lambda^{00}_{x,ii} & \Lambda^{02}_{x,ii}\\ 
\noalign{\smallskip}
\Lambda^{20}_{x,ii} & \Lambda^{22}_{x,ii}\\
\end{array}
\right), 
\label{vector_lambda_mat}
\end{eqnarray}
where $\Lambda^{KK^\prime}_{x,ii}$ are given by (see Equation~(\ref{vector_lambda}))
\begin{equation}
\Lambda^{KK^\prime}_{x,ii} = {1\over 2} \int_{-1}^{+1} 
\Psi^{KK^\prime}_0(\mu)\, \Lambda^{KK^\prime}_{x\mu,ii}\,{\rm d}\mu.
\label{vector_lambda_matelement}
\end{equation}
As shown by \citet{jtbrms99} for the CRD problem (see their Fig.~3), we 
also find for our PRD problem that $|\Lambda^{00}_{x,ii}| > 
|\Lambda^{22}_{x,ii}| \gg |\Lambda^{02}_{x,ii}| = |\Lambda^{20}_{x,ii}|$. 
We illustrate this fact in Fig.~\ref{lambda_element}, where the elements 
of the monochromatic lambda matrix for a semi-infinite model atmosphere 
is plotted versus the line center optical depth for two different frequencies 
$x=0$ (panel a) and $x=5$ (panel b). Clearly, for $x=0$ the elements of 
${\bf{\Lambda}}_{x,ii}$ are nearly identical to that of the corresponding CRD 
case \citep[compare our Fig.~\ref{lambda_element}a with Fig.~3 of][]{jtbrms99}. 
As the frequency increases toward the line wing the entire curve 
corresponding to all the elements of this matrix shifts toward higher 
optical depths (see Fig.~\ref{lambda_element}b). In other words, the depth at 
which $\Lambda^{KK^\prime}_{x,ii}$ reaches unity, $0.7$ and zero for 
$(K,K^\prime)=(0,0)$, $(2,2)$ and $(0,2)$, respectively, shifts toward 
larger optical depth. This behaviour can be understood by looking at the
explicit form of $\Lambda^{KK^\prime}_{x\mu,ii}$. In the case of a 
short-characteristic formal solver, $\Lambda^{KK^\prime}_{x\mu,ii} 
\approx \Psi_{x,O}(\mu)$, where $O$ is the depth point of interest (here 
$O=i$). The quantity $\Psi_{x,O}(\mu)$ is given by
\begin{equation}
\Psi_{x,O}(\mu) = w_0 - {(\Delta \tau_P-\Delta\tau_M)w_1 + w_2 \over 
\Delta \tau_P \Delta\tau_M},
\label{psixomu}
\end{equation}
with $w_0=1-\exp(-\Delta\tau_M)$, $w_1=w_0-\Delta\tau_M \exp(-\Delta\tau_M)$ 
and $w_2=2w_1-\Delta\tau^2_M \exp(-\Delta\tau_M)$. Clearly, when both 
$\Delta\tau_M$ and $\Delta\tau_P$ tend to infinity, $\Psi_{x,O}(\mu) \to 1$, 
and then it is easy to see from Equations~(\ref{vector_lambda_matelement}), 
(\ref{psikkprime0}) and (\ref{tk0_expression}) that 
$\Lambda^{KK^\prime}_{x,ii}$ saturate to their 
respective values mentioned 
above. As $x$ increases the optical depth at which $\Psi_{x,O}(\mu) \to 1$ 
shifts to larger optical depths, and hence the observed behaviour. Thus, 
saturation is reached when $\Psi_{x,O}(\mu) \to 1$. This is perhaps equivalent 
to saying that saturation is reached when the exponential in the kernel 
\citep[see Equation~(19) of][]{fauetal97} can be replaced reasonably well by a 
delta function in the grid interval around the optical depth $\tau_i$. It is 
worth noting that as $\Lambda^{KK^\prime}_{x,ii}$ depends on the optical 
depth grid, the finer the grid the slightly larger is the depth at which 
saturation is reached. 

Thus, following \citet{jtbrms99} we can also set 
$\Lambda^{22}_{x,ii}=\Lambda^{02}_{x,ii}=\Lambda^{20}_{x,ii}=0$, and still 
obtain a radiative transfer method with a convergence rate that is as good 
as that achieved by keeping the full ${\bf \Lambda}_{x,ii}$ given in 
Equations~(\ref{vector_lambda_mat}) 
and (\ref{vector_lambda_matelement}) \citep[see Fig.~4 of][]{jtbrms99}. 
The use of such a strategy leads to a decoupling 
of the equations for $(\delta S_{lx,i})^0_0$ and $(\delta S_{lx,i})^2_0$ 
as follows (see Equations~(\ref{vectordeltaslx_correction}), 
(\ref{nxxprime}), and (\ref{vector_lambda_mat}))\,: 
\begin{eqnarray}
&\!\!\!\!\!\!\!
(\delta S_{lx,i})^0_0 - \int_{-\infty}^{+\infty} (N_{xx^\prime})_{00} \,
p_{x^\prime}\, \Lambda^{00}_{x^\prime,ii}\, (\delta S_{lx^\prime,i})^0_0 \,
{\rm d}x^\prime\nonumber \\
&\!\!\!\!\!\!\!= \epsilon B + (\bar {J}^{\rm \,old}_{x,i})^0_0
- (S^{\rm old}_{lx,i})^0_0\,, 
\label{deltaslx00_trick}
\end{eqnarray}
\begin{equation}
(\delta S_{lx,i})^2_0 
= W_2\ (\bar{J}^{\rm \,old}_{x,i})^2_0 - (S^{\rm old}_{lx,i})^2_0\,, 
\label{deltaslx20_trick}
\end{equation}
where $(N_{xx^\prime})_{00}$ is the first diagonal element 
of ${\bf N}_{xx^\prime}$ given in Equation~(\ref{nxxprime}). 

In summary, the $(\delta S_{lx,i})^0_0$ correction is computed using a Jacobi, 
GS or SOR iteration, while the 
$(\delta S_{lx,i})^2_0$ correction is formally similar to the 
classical $\Lambda$-iteration. However, the important difference with respect 
to the classical $\Lambda$-iteration method is that the $(I_{x\mu,i})^0_0$ 
which enters the computation of $(J_{x,i})^2_0$ (see Equation~(\ref{vectorj})) 
is calculated with the improved $(S_{x,i})^0_0$ value that was obtained in the 
previous iterative step. 

Finally, we remark the following two important points\,: (1) As in the case of 
unpolarized transfer, in the polarized case we again find that as long as 
$\Gamma_{\rm E}/\Gamma_{\rm R} < 10$ we can approximate 
$g^{\rm III}_{xx^\prime}$ by $\phi_{x^\prime}$ in the core and neglect it in 
the wings. But as soon as $\Gamma_{\rm E}/\Gamma_{\rm R} \ge 10$, to get the 
converged solution we have to approximate $g^{\rm III}_{xx^\prime}$ by 
$\phi_{x^\prime}$ throughout the line profile for the line source vector 
correction computation. (2) All the above-mentioned iterative schemes are 
given for the DH redistribution matrix. It is not difficult to apply 
them for the hybrid approximation (see Equation~(\ref{hybrid_approx})). In this 
case we simply replace ${\bf N}_{xx^\prime}$ by $(1-\epsilon) 
g^{k}_{xx^\prime} {\bf E}$ and in 
Equations~(\ref{vector_line_source_corrections_crdcs}) and (\ref{vector_deltat}) 
set $\alpha=(1-\epsilon)$ and ${\bm{\mathcal B}}=(1-\epsilon){\bf E}$. 
Furthermore, when applying CRD-CS for $k={\rm I,III}$ we do the same 
approximations as we did for unpolarized case (see \S~3.1.2.). 

\section{The true error for $(S_x)^2_0$}
\begin{figure*}
%\centering   
%\includegraphics[height=16cm,width=16cm]{tecerc_pol.eps}
\plotone{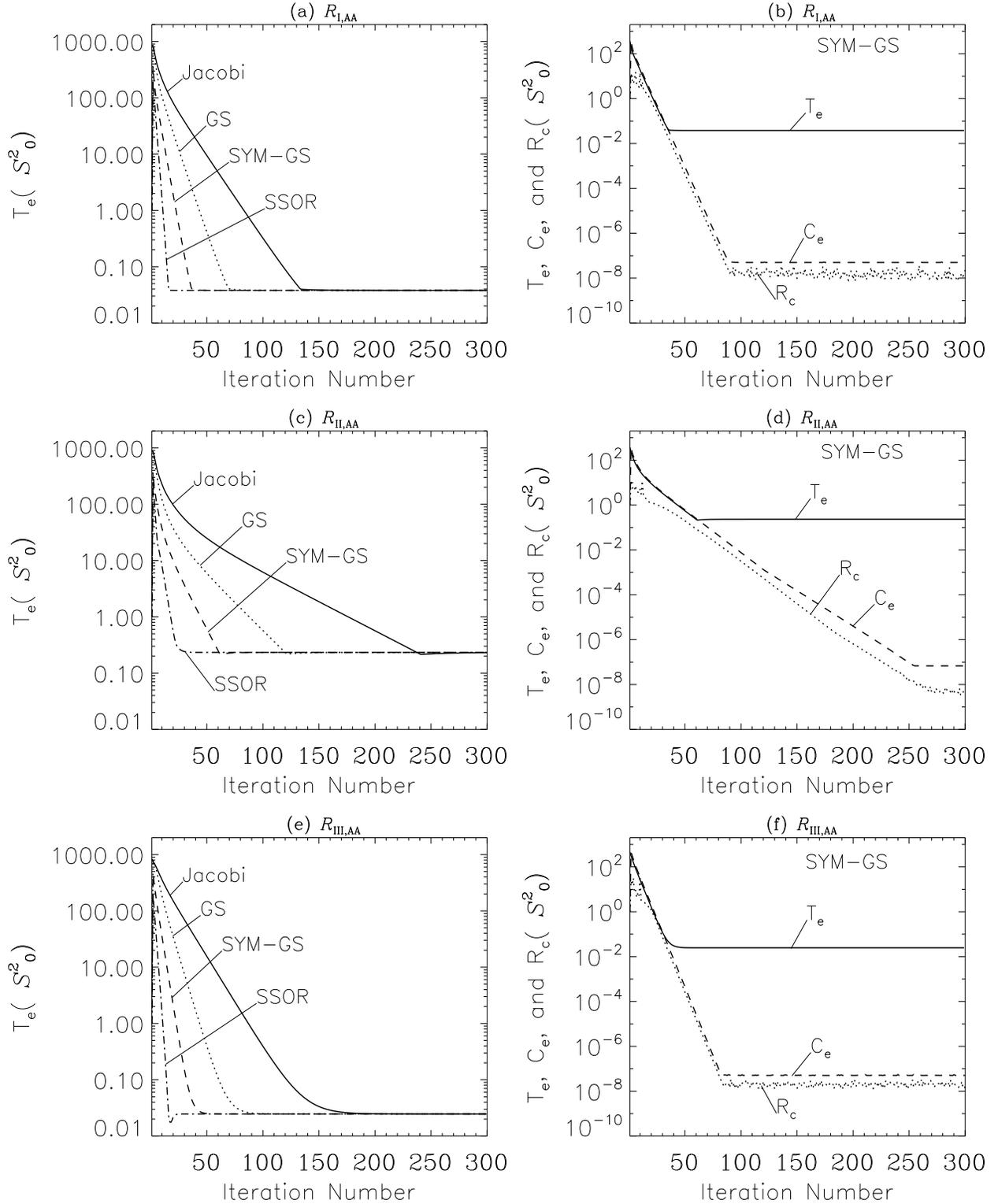}
\caption{$T_e$, $C_e$ and $R_c$ for $(S_x)^2_0$. 
Convergence properties of the various iterative schemes. 
Left panels\,: solid line (Jacobi), 
dotted line (GS), dashed line (SYM-GS), dot-dashed line (SSOR, 
$\omega_{\rm opt}=1.6$). Right panels\,: solid line ($T_e$), 
dotted line ($R_c$), and dashed line ($C_e$), computed 
using the SYM-GS iterative scheme. For type II and type III 
redistribution the damping parameter $a=10^{-3}$. A spatial grid with 9 points 
per decade ($\Delta Z=0.25$) is used. 
}
\label{rctece_pol}
\end{figure*}

\begin{figure}%[ht!]
%\centering  
%\includegraphics[height=9.5cm,width=7.5cm]{trueerror_dh_gebyr.eps}
\epsscale{0.8}
\plotone{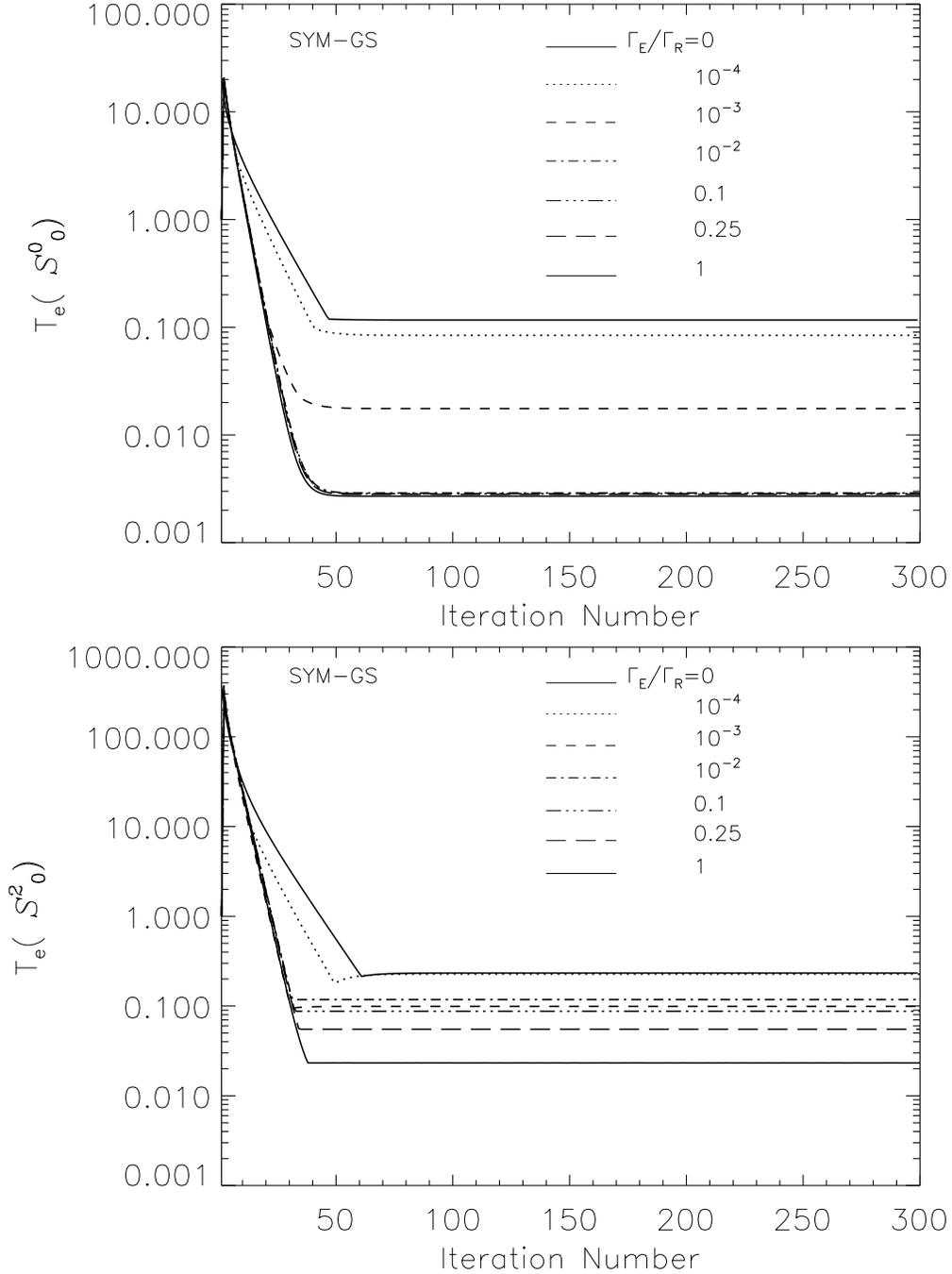}
\caption{True error for the DH redistribution function problem. 
Effect of the elastic collision parameter $\Gamma_{\rm E}/\Gamma_{\rm R}$. 
The non-LTE parameter $\epsilon=10^{-4}$, the damping parameter 
$a=10^{-3}$, $r=0$ (pure line case), $D^{(2)}=0.5\,\Gamma_{\rm E}$, and 
depth grid spacing $\Delta Z=0.25$. The topmost solid 
line\,: $\Gamma_{\rm E}/\Gamma_{\rm R}=0$, dotted line\,: 
$\Gamma_{\rm E}/\Gamma_{\rm R}=10^{-4}$, dashed line\,: 
$\Gamma_{\rm E}/\Gamma_{\rm R}=10^{-3}$, 
dot-dashed line\,: $\Gamma_{\rm E}/\Gamma_{\rm R}=10^{-2}$, 
dash-triple-dotted line\,: $\Gamma_{\rm E}/\Gamma_{\rm R}=0.1$, 
long-dashed line\,:
$\Gamma_{\rm E}/\Gamma_{\rm R}=0.25$, and the bottom solid line\,: 
$\Gamma_{\rm E}/\Gamma_{\rm R}=1$. 
}
\label{te_dh_gebyr}
\end{figure}

In \S~4 we presented a detailed study of the true error for the 
unpolarized PRD source function. In this section we present similar studies 
for $(S_x)^2_0$. We note that the behaviour of $T_e$, $C_e$ and $R_c$ 
presented in \S~4 for the unpolarized source function is similar 
for $(S_x)^0_0$. Therefore, we consider only $(S_x)^2_0$ 
here. The results are presented in Figs.~\ref{rctece_pol} and 
\ref{te_dh_gebyr}. 

The definition of $T_e$, $C_e$ and $R_c$ for $(S_x)^2_0$ is also given by 
Equations~(\ref{mrc})--(\ref{te}), but with $S_{lx}$ replaced by $(S_x)^2_0$. 
%and $S_{lx}$ in the denominator of those equations replaced by $|(S_x)^2_0|$, 
%as the latter is a sign changing quantity. 
However, $(S_x)^2_0$ is a sign changing quantity. Thus, in the denominator 
of Equations~(\ref{mrc})--(\ref{te}), we need to replace $S_{lx}$ 
by $|(S_x)^2_0|$. 
Furthermore, it is well known that 
in general $(S_x)^2_0$ is two orders of magnitude smaller than $(S_x)^0_0$. 
Therefore, we find that in our PRD case the maximum relative change $R_c$ 
for $(S_x)^2_0$ reaches approximately a minimum value of $10^{-8}$ and then 
starts to fluctuate around it. Thus, to find a fully converged solution on 
a given grid resolution level $g$, we iterate until $R_c < 10^{-8}$.  
We remark that we adopt the same method as described 
in \S~4 to find the true error as well as the convergence error for 
$(S_x)^2_0$. 

As in \S~4, here we consider the hybrid approximation with 
$R_{\rm I,II,III,AA}$ redistribution functions and the DH redistribution 
matrix. Again, we consider a semi-infinite atmosphere with the lower boundary 
condition $(I_{x\mu})^K_0=B\,\delta_{K0}$, and the upper boundary condition 
$(I_{x\mu})^K_0=0$ for $K=0,2$. A depth grid of 9 points per decade (i.e., 
$\Delta Z=0.25$) and a Gaussian quadrature with 5 $\mu$-values $[0<\mu<1]$ 
are used. The frequency grid used is exactly the same as that chosen for the 
unpolarized case. Other parameters are $\epsilon=10^{-4}$, $r=0$, 
$B=1$, and $a=10^{-3}$ for type II and type III redistribution, unless stated 
otherwise. We initialize all the three iterative schemes discussed in this 
paper by the LTE solution\,: $(S_{lx})^K_0=B\,\delta_{K0}$. It is 
worthwhile to note that if the initial or starting solution is other than the 
above-mentioned LTE solution, then the true error that one obtains for a given 
grid resolution (after reaching the plateau region of the $T_e$ curve) remains 
the same. However, the path followed to reach that $T_e$ is different. 

We find that the true error for $(S_x)^2_0$ is in general larger by one order 
of magnitude compared to that for $(S_x)^0_0$. Fig.~\ref{rctece_pol} 
shows the $T_e$, $C_e$ and $R_c$ for $(S_x)^2_0$ with the hybrid approximation 
and $R_{\rm I,II,III,AA}$ redistribution functions. As in the unpolarized 
case, the true error for $(S_x)^2_0$ is the largest for the $R_{\rm II,AA}$ 
redistribution function case. It is about 23\,\%. However, as discussed 
for the unpolarized case, addition of the continuum or inclusion of elastic 
collisions through the use of the DH redistribution matrix improves the true 
error significantly. For example, the true error for $(S_x)^2_0$ is 
approximately 2.3\,\% for $\Gamma_{\rm E}/\Gamma_{\rm R}=1$ (see the bottom 
solid line in Fig.~\ref{te_dh_gebyr}) without continuum, and it is 1\,\% for 
$\Gamma_{\rm E}/\Gamma_{\rm R}=0$ and $r=10^{-4}$ (figure not shown). 

\section{Conclusions}

In this paper we have shown how to solve efficiently and accurately 
the non-LTE resonance line formation problem in stellar atmospheres, 
taking into account PRD effects with and without scattering polarization. 
To this end, we have generalized the Gauss-Seidel (GS) and 
Successive Over-Relaxation (SOR) radiative transfer methods that 
\citet{jtbandfb95} and \citet{jtbrms99} developed for solving 
unpolarized and polarized CRD problems, respectively. These iterative 
methods are based on the concept of operator splitting. As in the CRD case, 
we find that these methods are superior to the Jacobi-based ALI method. 
Quantitatively, the symmetric GS method (SYM-GS) is 4 times faster than Jacobi, 
while the symmetric SOR method (SSOR) is about 10 times faster without 
the need of refining the choice of the $\omega$-parameter. 
We emphasize that our implementation of 
these highly convergent radiative transfer methods do not require 
neither the construction nor the inversion of any non-local 
$\Lambda$-operator, so that the computing time per iteration is similar to that of the 
Jacobi method. Therefore, these GS-based methods are suitable also for the solution of 
non-LTE problems in three-dimensional model atmospheres. 

For the unpolarized PRD problem, we have considered the case of 
pure Doppler redistribution (type I), Doppler, natural and collisionally 
broadened type III redistribution, Doppler and naturally broadened type 
II redistribution, and a combined case of type II and type III redistribution. 
For the PRD problem of resonance line polarization we have considered both the 
hybrid approximation with angle-averaged type I, II and III redistribution 
functions and the general redistribution matrix of \citet{dh88} that properly 
takes into account the elastic and depolarizing collisions. The methods 
we have developed here can be used also for solving the resonance line 
polarization problem in the presence of a magnetic field that does not 
break the axial symmetry of the problem. For the case 
of a weak magnetic field with a given strength, inclination and 
azimuth at each spatial grid point the corresponding redistribution 
matrices are substantially more complicated \citep[e.g.,][]{bom97}, 
but the generalization of the same GS-based methods is straightforward 
in spite of the fact that the number of unknowns is three times larger.

%We present a detailed study of the true error of the converged PRD solutions in spatial grids of increasing resolution. We find that it depends not only on the resolution of the grid and on the accuracy of the formal solver, but also on the choice of the redistribution function. For a given grid resolution, the largest truncation error is found for the type II redistribution case. This is due to the coherent nature of the $R_{\rm II,AA}$ function in the wings, which cannot be easily thermalized without the presence of a background continuum. However, the addition of the continuum opacity or elastic collisions decreases the true error to sufficiently low values for all practical purposes.

Finally, we emphasize that the PRD radiative transfer problem we have 
considered here is that of a two-level model atom without the possibility of 
lower-level polarization, which implies that it is assumed that only the emission 
term of the transfer equation contributes to scattering polarization. Fortunately, 
there are several diagnostically important resonance lines for which this 
two-level atom approximation is probably suitable (e.g., the $k$ line of Mg {\sc ii}). 
In forthcoming papers we will show how the application of 
the computer programs described here allow us to gain physical insight and to 
make predictions on the $Q/I$ shapes produced by PRD effects.

\acknowledgments
Financial support by the Spanish Ministry of Science and Innovation 
through projects AYA2007-63881 (Solar Magnetism and High-Precision 
Spectropolarimetry) and CONSOLIDER INGENIO CSD2009-00038 (Molecular 
Astrophysics: The Herschel and Alma Era) is gratefully acknowledged.
We are also grateful to the referee for carefully reviewing our paper.

\eject
%%%%%%%%%%%%%%%%%%%%%%%%%%%%%%%%%%%%%%%%%%%%%%%%%%%%%%%%%%%%%%%%%%table1
\begin{deluxetable}{ccc}
\tablewidth{0pt}
\tablecaption{The true error in the case of type II redistribution for 
different resolutions of the depth grid. For all cases, 
$\epsilon=10^{-4}$, $a=10^{-3}$ and $r=0$.
\label{r2_te_depth}}
\tablehead{
\colhead{$\Delta Z$} & \colhead{Number of points} &
\colhead{$T_e$}\\
\colhead{} & \colhead{per decade} &
\colhead{}
}
\startdata
0.5  & 4.5 & 0.2571 \\
0.25 & 9 & 0.1244 \\
0.125 & 18 & 0.0638 \\
0.1 & 23 & 0.0533  \\
0.04 & 28.75 & 0.0238 \\
0.02 & 57.5 & 0.0130 \\
0.01 & 115  & 0.0069 \\
\enddata
\end{deluxetable}
%%%%%%%%%%%%%%%%%%%%%%%%%%%%%%%%%%%%%%%%%%%%%%%%%%%%%%%%%%%%%%%%%%table1
%%%%%%%%%%%%%%%%%%%%%%%%%%%%%%%%%%%%%%%%%%%%%%%%%%%%%%%%%%%%%%%%%%fig1
%%%%%%%%%%%%%%%%%%%%%%%%%%%%%%%%%%%%%%%%%%%%%%%%%%%%%%%%%%%%%%%%%%fig1
%%%%%%%%%%%%%%%%%%%%%%%%%%%%%%%%%%%%%%%%%%%%%%%%%%%%%%%%%%%%%%%%%%fig2
%%%%%%%%%%%%%%%%%%%%%%%%%%%%%%%%%%%%%%%%%%%%%%%%%%%%%%%%%%%%%%%%%%fig2
%%%%%%%%%%%%%%%%%%%%%%%%%%%%%%%%%%%%%%%%%%%%%%%%%%%%%%%%%%%%%%%%%%fig3
%%%%%%%%%%%%%%%%%%%%%%%%%%%%%%%%%%%%%%%%%%%%%%%%%%%%%%%%%%%%%%%%%%fig3
%%%%%%%%%%%%%%%%%%%%%%%%%%%%%%%%%%%%%%%%%%%%%%%%%%%%%%%%%%%%%%%%%%fig4
%%%%%%%%%%%%%%%%%%%%%%%%%%%%%%%%%%%%%%%%%%%%%%%%%%%%%%%%%%%%%%%%%%fig4
%%%%%%%%%%%%%%%%%%%%%%%%%%%%%%%%%%%%%%%%%%%%%%%%%%%%%%%%%%%%%%%%%%fig5
%%%%%%%%%%%%%%%%%%%%%%%%%%%%%%%%%%%%%%%%%%%%%%%%%%%%%%%%%%%%%%%%%%fig5
%%%%%%%%%%%%%%%%%%%%%%%%%%%%%%%%%%%%%%%%%%%%%%%%%%%%%%%%%%%%%%%%%%fig6
%%%%%%%%%%%%%%%%%%%%%%%%%%%%%%%%%%%%%%%%%%%%%%%%%%%%%%%%%%%%%%%%%%fig6
%%%%%%%%%%%%%%%%%%%%%%%%%%%%%%%%%%%%%%%%%%%%%%%%%%%%%%%%%%%%%%%%%%fig7
%%%%%%%%%%%%%%%%%%%%%%%%%%%%%%%%%%%%%%%%%%%%%%%%%%%%%%%%%%%%%%%%%%fig7
%%%%%%%%%%%%%%%%%%%%%%%%%%%%%%%%%%%%%%%%%%%%%%%%%%%%%%%%%%%%%%%%%%fig8
%%%%%%%%%%%%%%%%%%%%%%%%%%%%%%%%%%%%%%%%%%%%%%%%%%%%%%%%%%%%%%%%%%fig8
%%%%%%%%%%%%%%%%%%%%%%%%%%%%%%%%%%%%%%%%%%%%%%%%%%%%%%%%%%%%%%%%%%fig9
%%%%%%%%%%%%%%%%%%%%%%%%%%%%%%%%%%%%%%%%%%%%%%%%%%%%%%%%%%%%%%%%%%fig9

\begin{thebibliography}{}
\bibitem[Anusha et al.(2009)]{lsaetal09}
Anusha, L.~S., Nagendra, K.~N., Paletou, F., \& L\'eger, L. 2009, \apj, 704, 661
\bibitem[Asensio Ramos \& Trujillo Bueno(2006)]{araandjtb06}
Asensio Ramos, A., \& Trujillo Bueno, J. 2006, in EAS Publication Series 18, Radiative 
Transfer and Applications to Very Large Telescopes, ed. Ph. Stee (EAS, EDP Sciences) 25
\bibitem[Auer(1987)]{aue87}
Auer, L. 1987, in Numerical Radiative Transfer, ed. W. Kalkofen 
(Cambridge: Cambridge Univ. Press), 101
\bibitem[Auer(1991)]{aue91}
Auer, L. 1991, in Stellar Atmospheres: Beyond Classical Models, ed. 
L. Crivellari, I. Hubeny, \& D. G. Hummer (Dordrecht: Kluwer), 9
\bibitem[Auer et al.(1994)Auer, Fabiani Bendicho \& Trujillo Bueno]{aueetal94}
Auer, L., Fabiani Bendicho, P., \& Trujillo Bueno, J. 1994, \aap, 292, 
599
\bibitem[Auer \& Paletou(1994)]{aueandpal94}
Auer, L., \& Paletou, F. 1994, \aap, 285, 675
\bibitem[Belluzzi \& Landi Degl'Innocenti(2009)]{lbandeld09}
Belluzzi, L., \& Landi Degl'Innocenti, E. 2009, \aap, 495, 577
\bibitem[Bommier(1997)]{bom97}
Bommier, V. 1997, \aap, 328, 726
\bibitem[Cannon(1973)]{can73}
Cannon, C. J. 1973, \apj, 185, 621
\bibitem[Cannon(1985)]{can85}
Cannon, C.~J. 1985, The transfer of spectral line radiation 
(Cambridge: Cambridge University Press)
%\bibitem[Cannon \& Vardavas(1974)]{canandvar74}
%Cannon, C. J., \& Vardavas, I. M. 1974, \aap, 32, 85  
\bibitem[Castor(2004)]{cas04}
Castor, J. 2004, Radiation Hydrodynamics (Cambridge University Press)
\bibitem[Chandrasekhar(1950)]{chandra50}
Chandrasekhar, S. 1950, Radiative transfer (Oxford: Clarendon Press)
\bibitem[Chevallier et al.(2003)]{chevaletal03}
Chevallier, L., Paletou, F., \& Rutily, B. 2003, \aap, 411, 221
\bibitem[Domke \& Hubeny(1988)]{dh88}
Domke, H., \& Hubeny, I. 1988, \apj, 334, 527
\bibitem[Fabiani Bendicho \& Trujillo Bueno(1999)]{fbpandjtb99}
Fabiani Bendicho, P., \& Trujillo Bueno, J. 1999, in Solar Polarization, ed. K. N. 
Nagendra, \& J. O. Stenflo (Boston: Kluwer), 219
\bibitem[Fabiani Bendicho et al.(1997)Fabiani Bendicho, Trujillo Bueno \& Auer]{fbpetal97}
Fabiani Bendicho, P., Trujillo Bueno, J., \& Auer, L. 1997, \aap, 324, 161
\bibitem[Faurobert(1987)]{fau87}
Faurobert, M. 1987, \aap, 178, 269
\bibitem[Faurobert(1988)]{fau88}
Faurobert, M. 1988, \aap, 194, 268
\bibitem[Faurobert-Scholl(1992)]{fau92}
Faurobert-Scholl, M., 1992, \aap, 258, 521
\bibitem[Faurobert-Scholl(1993)]{fau93}
Faurobert-Scholl, M., 1993, \aap, 268, 765
\bibitem[Faurobert-Scholl et al.(1997)]{fauetal97}
Faurobert-Scholl, M., Frisch, H., \& Nagendra, K.~N. 1997, \aap, 322, 896
\bibitem[Feautrier(1964)]{fea64}
Feautrier, P. 1964, C.~R. Acad. Sci. Paris, 258, 3189
\bibitem[Fluri et al.(2003)]{fluetal03}
Fluri, D.~M., Nagendra, K.~N., \& Frisch, H. 2003, 
\aap, 400, 303
\bibitem[Frisch(1980)]{hf80}
Frisch, H. 1980, \aap, 83, 166
\bibitem[Frisch(1988)]{hf88}
Frisch, H. 1988, in Radiation in moving gaseous media, ed.  Y. Chmielewski, 
\& T. Lanz (Switzerland: Geneva Observatory), 337
\bibitem[Frisch(2007)]{hf07}
Frisch, H. 2007, \aap, 476, 665
\bibitem[Frisch et al.(2009)]{hfetal09}
Frisch, H., Anusha, L.~S., Sampoorna, M., \& Nagendra, K.~N. 2009, \aap, 501, 
335
\bibitem[Gandorfer(2000)]{gan00}
Gandorfer, A. 2000, The Second Solar Spectrum, Vol I\,: 4625\,\AA\ to
6995\,\AA\ (Zurich\,: vdf Hochschulverlag)
\bibitem[Gandorfer(2002)]{gan02}
Gandorfer, A. 2002, The Second Solar Spectrum, Vol II\,: 3910\,\AA\ to
4630\,\AA\ (Zurich\,: vdf Hochschulverlag)
\bibitem[Gandorfer(2005)]{gan05}
Gandorfer, A. 2005, The Second Solar Spectrum, Vol III\,: 3160\,\AA\ to
3915\,\AA\ (Zurich\,: vdf Hochschulverlag)
\bibitem[Heinzel(1981)]{hei81}
Heinzel, P. 1981, \jqsrt, 25, 483
\bibitem[Hubeny(2003)]{hubeny03}
Hubeny, I. 2003, in ASP Conf. Ser. 288, Stellar Atmosphere Modeling, ed.
I. Hubeny, D. Mihalas, \& K. Werner (San Francisco: ASP), 17
\bibitem[Hummer(1962)]{hum62}
Hummer, D.~G. 1962, \mnras, 125, 21
%\bibitem[Hummer(1969)]{hum69}
%Hummer, D. G. 1969, \mnras, 145, 95
\bibitem[Ivanov(1995)]{iva95}
Ivanov, V.~V. 1995, \aap, 303, 609
\bibitem[Klein et al.(1989)]{kleetal89}
Klein, R. I., Castor, J. I., Greenbaum, A., Taylor, D., \& Dykema, P. G. 
1989, \jqsrt, 41, 199
\bibitem[Kunasz \& Auer(1988)]{kunandaue88}
Kunasz, P., \& Auer, L. H. 1988, \jqsrt, 39, 67
\bibitem[Landi Degl'Innocenti(1984)]{lan84}
Landi Degl'Innocenti, E. 1984, \solphys, 91, 1
\bibitem[Landi Degl'Innocenti \& Landolfi(2004)]{ll04}
Landi Degl'Innocenti, E., \& Landolfi, M. 2004, Polarization in spectral 
lines (Dordrecht: Kluwer) 
\bibitem[Manso Sainz \& Trujillo Bueno(1999)]{rmsjtb99}
Manso Sainz, R., \& Trujillo Bueno, J. 1999, in Solar Polarization, ed. K. N. 
Nagendra, \& J. O. Stenflo (Boston: Kluwer), 143
\bibitem[Manso Sainz \& Trujillo Bueno(2003)]{rmsjtb03}
Manso Sainz, R., \& Trujillo Bueno, J. 2003, in ASP Conf. Ser. 307, 
Solar Polarization, ed. J. Trujillo Bueno, \& J. S\'anchez Almeida 
(San Francisco: ASP), 251
%\bibitem[McKenna(1980)]{mck80}
%McKenna, S.~J. 1980, \apj, 242, 283
\bibitem[McKenna(1984)]{mck84}
McKenna, S.~J. 1984, \apss, 106, 283
\bibitem[Mihalas(1978)]{mih78}
Mihalas, D. 1978, Stellar atmosphere (2nd ed.; San Francisco: Freeman)
\bibitem[Nagendra(1986)]{knn86}
Nagendra, K.~N. 1986, PhDT, 
Radiative transfer with Stokes vector (Bangalore: Bangalore University)
\bibitem[Nagendra(1988)]{knn88}
Nagendra, K.~N. 1988, \apj, 335, 269
\bibitem[Nagendra(1989)]{knn89}
Nagendra, K.~N. 1989, \apss, 154, 119
\bibitem[Nagendra(1994)]{knn94}
Nagendra, K.~N. 1994, \apj, 432, 274
\bibitem[Nagendra(1995)]{knn95}
Nagendra, K.~N. 1995, \apss, 274, 523
\bibitem[Nagendra(2003)]{knn03}
Nagendra, K.~N. 2003, in ASP Conf. Ser. 288, Stellar Atmosphere Modeling, ed.
I. Hubeny, D. Mihalas, \& K. Werner (San Francisco: ASP), 583
\bibitem[Nagendra et al.(2009)]{knnetal09}
Nagendra, K.~N., Anusha, L.~S., \& Sampoorna, M. 2009, \memsai, 80, 678
\bibitem[Nagendra et al.(1998)]{knnetal98}
Nagendra, K.~N., Frisch, H., \& Faurobert-Scholl, M. 1998, \aap, 332, 610
\bibitem[Nagendra et al.(1999)]{knnetal99}
Nagendra, K. N., Paletou, F., Frisch, H., \& Faurobert-Scholl, M. 1999, in Solar 
Polarization, ed. K.~N. Nagendra, \& J.~O. Stenflo (Boston: Kluwer), 127
\bibitem[Nagendra \& Sampoorna(2009)]{knnms09}
Nagendra, K.~N., \& Sampoorna, M. 2009, in ASP Conf. Ser. 405, Solar 
Polarization 5, ed.
S.~V. Berdyugina, K.~N. Nagendra, \& R. Ramelli (San Francisco: ASP), 261
\bibitem[Olson, Auer \& Buchler(1986)]{olsetal86}
Olson, G.~L., Auer, L.~H., \& Buchler, J.~R. 1986, \jqsrt, 
35, 431
\bibitem[Omont et al.(1972)]{osc72}
Omont, A., Smith, E.~W., \& Cooper, J. 1972, \apj, 175, 185
\bibitem[Oxenius \& Simonneau(1994)]{os94}
Oxenius, J., \& Simonneau, E. 1994, Annals of Physics, 234, 60
\bibitem[Paletou \& Anterrieu(2009)]{pfanda09}
Paletou, F., \& Anterrieu, E. 2009, \aap, 507, 1815
\bibitem[Paletou \& Auer(1995)]{palandaue95}
Paletou, F., \& Auer, L.~H. 1995, \aap, 297, 771 (PA95)
\bibitem[Paletou \& Faurobert-Scholl(1997)]{fpmf97}
Paletou, F., \& Faurobert-Scholl, M. 1997, \aap, 328, 343
\bibitem[Rees \& Saliba(1982)]{reeandsal82}
Rees, D.~E., \& Saliba, G.~J. 1982, \aap, 115, 1  
%\bibitem[Sampoorna et al.(2007)]{sametal07}
%Sampoorna, M., Nagendra, K. N., \& Stenflo, J. O. 2007, \apj, 663, 625
\bibitem[Sampoorna et al.(2008)]{sametal08}
Sampoorna, M., Nagendra, K. N., \& Frisch, H. 2008, \jqsrt, 109, 2349 
\bibitem[Scharmer(1983)]{sch83}
Scharmer, G. B. 1983, \aap, 117, 83
\bibitem[Stenflo(1994)]{ste94}
Stenflo, J. O. 1994, Solar magnetic fields - Polarized radiation diagnostics 
(Dordrecht: Kluwer)
\bibitem[Stenflo \&\ Keller(1997)]{josck97}
Stenflo, J.~O., \& Keller, C.~U. 1997, \aap, 321, 927
\bibitem[Trujillo Bueno(1999)]{jtb99}
Trujillo Bueno, J. 1999, in Solar Polarization, ed. K.~N. Nagendra,
\& J.~O. Stenflo (Boston: Kluwer), 73
\bibitem[Trujillo Bueno(2003)]{jtb03}
Trujillo Bueno, J. 2003, in ASP Conf. Ser. 288, 
Stellar atmosphere modeling, eds. I. Hubeny, D. Mihalas, \& K.
Werner (San Francisco: ASP), 551
\bibitem[Trujillo Bueno(2009)]{jtb09}
Trujillo Bueno, J. 2009, in ASP Conf. Ser. 405, Solar Polarization 5, ed.
S.~V. Berdyugina, K. N. Nagendra, \& R. Ramelli (San Francisco: ASP), 65
\bibitem[Trujillo Bueno \& Fabiani Bendicho(1995)]{jtbandfb95}
Trujillo Bueno, J., \& Fabiani Bendicho, P. 1995, \apj, 455, 646 (TF95)
\bibitem[Trujillo Bueno \& Landi Degl'Innocenti(1996)]{jtbandeld96}
Trujillo Bueno, J., \& Landi Degl'Innocenti, E. 1996, \solphys, 164, 135
\bibitem[Trujillo Bueno \& Manso Sainz(1999)]{jtbrms99}
Trujillo Bueno, J., \& Manso Sainz, R. 1999, \apj, 516, 436 
\bibitem[Uitenbroek(2001)]{rh01}
Uitenbroek, H. 2001, \apj, 557, 389
%\bibitem[Vardavas(1976a)]{var76a}
%Vardavas, I. M. 1976a, \jqsrt, 16, 1
%\bibitem[Vardavas(1976b)]{var76b}
%Vardavas, I. M. 1976b, \jqsrt, 16, 715
\bibitem[Vardavas \& Cannon(1976)]{varandcan76}
Vardavas, I. M., \& Cannon, C. J. 1976, \aap, 53, 107
\bibitem[Vinsome(1976)]{vin76}
Vinsome, P. K. W. 1976, in Proc. 4th Symp. on Numerical Simulation of 
Reservoir Performance of SPE of AIME, 140
%\bibitem[Wallace \& Yelle(1989)]{walandyel89}
%Wallace, L., \& Yelle, R.~V. 1989, \apj, 346, 489
\end{thebibliography}
\end{document}